\documentclass[12pt]{article}

\usepackage{amsmath}
\usepackage{cite}
\usepackage[xy,curve,matrix,oldcm,debug]{qsymbols}
\usepackage{hhline}

\def\nn{\nonumber}

\numberwithin{equation}{section}

\title{Lagrangian formulation of the massive\\ higher spin $N=1$ supermultiplets
 in $AdS_4$ space}

\author{I.L. Buchbinder${}^{ab}$\thanks{joseph@tspu.edu.ru},
M.V. Khabarov${}^{cd}$\thanks{maksim.khabarov@ihep.ru}, T.V.
Snegirev${}^{ae}$\thanks{snegirev@tspu.edu.ru}, Yu.M.
Zinoviev${}^{cd}$\thanks{Yurii.Zinoviev@ihep.ru}
\\[0.5cm]
\it{\small ${}^a$Department of Theoretical Physics, Tomsk State
Pedagogical University,}\\
\it{\small Tomsk, 634061, Russia}\\
\it{\small ${}^b$National Research Tomsk State University, Tomsk
634050, Russia}\\
\it{\small ${}^c$Institute for High Energy Physics of National
Research Center "Kurchatov Institute"} \\
\it{\small Protvino, Moscow Region, 142281, Russia} \\
\it\small{ ${}^d$Moscow Institute of Physics and Technology (State
University),} \\
\it{\small Dolgoprudny, Moscow Region, 141701, Russia}\\
\it{\small ${}^e$National Research Tomsk Polytechnic University,
Tomsk 634050, Russia}}

\date{}

\begin{document}

\maketitle

\begin{abstract}
We give an explicit component Lagrangian construction of massive
higher spin on-shell $N=1$ supermultiplets in four-dimensional
Anti-de Sitter space $AdS_4$. We use a frame-like gauge invariant
description of massive higher spin bosonic and fermionic fields. For
the two types of the supermultiplets (with integer and half-integer
superspins) each one containing two massive bosonic and two massive
fermionic fields we derive the supertransformations leaving the sum
of four their free Lagrangians invariant such that the algebra of
these supertransformations is closed on-shell.
\end{abstract}

\thispagestyle{empty}
\newpage
\setcounter{page}{1}
\tableofcontents\pagebreak

\section{Introduction}
The higher spin theory (see e.g. the reviews \cite{Vas04}
\cite{Be04},\cite{BBS10}) has attracted significant interest for a
long time and for many reasons. On the one hand, the theory of
massless higher spin fields is a maximal extension of the Yang-Mills
gauge theories and gravity including all spin fields. On the other
hand, it is closely related to superstring theory which involves an
infinite tower of higher spin massive fields. In principle, the
higher spin field theory can provide the possibility to study some
aspects of string theory in the framework of the field theory. It is
also worth pointing out that the construction of Lagrangian
formulations for the higher spin field models is extremely
interesting itself since it allows to reveal the new unexpected
properties to relativistic field theory in general.

Beginning with work \cite{FV} it became clear that the nonlinear
massless higher spin  theory can only be realized in $AdS$ space
with non-zero curvature. This raises the interest in studying the
various aspects of field theory in $AdS$ space in the context of
$AdS/CFT$-correspondence. Taking into account that the low-energy
limit of superstring theory should lead to supersymmetric field
theory we face the problem of constructing the supersymmetric
massive higher spin models in the $AdS$ space. It is expected that
the supersymmetry can be an essential ingredient of the consistent
theory of all the fundamental interactions including quantum
gravity. It is possible that such a theory should also involve the
massless and/or massive higher spin fields. This paper is devoted to
developing the $N=1$ supersymmetric Lagrangian formulation of free
massive higher spin models in $AdS$ space in the framework of
on-shell component formalism.

In supersymmetric theories the massless or massive fields are
combined into the corresponding supermultiplets. In the case of free
field models containing the different spin fields, it is natural to
expect that the Lagrangian should be the sum of the Lagrangians for
each concrete spin field. To provide an explicit Lagrangian
realization of the free  supermultiplet one has to find
supertransformations leaving the free Lagrangians invariant and show
that the algebra of these supertransformations is closed at least
on-shell. In the case of the $N=1$ supersymmetry the massless higher
superspin-$s$ supermultiplets consist of the  two massless fields
with spins $(s,s+1/2)$. The task of constructing
supertransformations for such supermultiplets in four dimensional
flat space was completely solved in the metric-like formulation
\cite{Cur79} and soon in the frame-like one \cite{Vas80}. In both
cases the supertransformations have a simple enough structure and
are determined uniquely by the invariance of the sum of the
Lagrangians for two free massless fields with spins $s$ and $s+1/2$.
Note that such a requirement allows to find only on-shell
supersymmetry when supertransformations are closed on the equations
of motion. In order to find off-shell supertransformations, it is
necessary to introduce the corresponding auxiliary fields.

A natural procedure to construct off-shell $N=1$ supersymmetric
Lagrangian models is realized in terms of $N=1$ superspace and
superfields (see e.g. \cite{BK98}), where all the auxiliary fields
providing closure of the superalgebra are automatically obtained. In
the framework of superfield formulation the $N=1$ supersymmetric
massless higher spin models were constructed in the pioneer papers
\cite{KSP93,KS93}. Later, on the basis of these results, $N=1$
supersymmetric massless higher spin models were generalized for
$AdS_4$ space \cite{KS94}, \cite{GKS}\footnote{Application of this
formulation for quantization of the $N=1$ higher spin superfield
model in $AdS_4$ space was considered in \cite{BKS}. The superfield
approach was recently applied for construction of the higher spin
supercurrents \cite{HK-1}, \cite{HK-2}, \cite{BHK}. \cite{BGK-1},
\cite{BGK-2}, \cite{BGK-3}, \cite{K}.}. In both cases the
constructed superfield models, after eliminating the auxiliary
fields, reduce to the sum of spin-$s$ and spin-$(s+1/2)$
(Fang)-Fronsdal Lagrangians \cite{Frons78,FF78} in flat or $AdS$
spaces. The generalization for ${\cal N}=2$ massless higher spin
supermultiplets was given in \cite{GKS}, \cite{GKS96a}.

There are much fewer results in the case of supersymmetric massive
higher spin models even in the on-shell formalism, the reason being
that when moving from the massless component formulation to the
massive one very complicated higher derivative corrections must be
introduced to the supertransformations. Moreover the higher the spin
of the fields entering a supermultiplet the higher the number of
derivatives one has to consider. The problem of the supersymmetric
description of the massive higher spin supermultiplet was only
explicitly resolved in 2007 for the case of $N=1$ on-shell 4D
Poincare superalgebra \cite{Zin07a} (see also
\cite{Zin02,Zin07,OT16}) using the gauge invariant formulation for
the massive higher spin fields
\cite{KZ97,Zin01,Met06}\footnote{Another gauge invariant approach to
Lagrangian formulation of massive higher spin fields is given on the
base of BRST construction \cite{BKr05}, \cite{BKrL}, \cite{BKRT},
\cite{BKrR}.}. In such a formalism the description for the massive
field is obtained in terms of the appropriately chosen set of the
massless ones. It is assumed that the Lagrangian for massive higher
spin supermultiplets is constructed as a sum of the corresponding
Lagrangians for massless fields deformed by massive terms. However,
it appeared \cite{Zin07a} that to realize such a program one has to
use massless supermultiplets containing four fields
$(k-1/2,k,k',k+1/2)$ as the building blocks, where two bosonic
fields with equal spins have opposite parities, and this prevents us
from separating them into the usual massless pairs. In \cite{Zin07a}
it was shown that to obtain the massive deformation it is enough to
add the non-derivative corrections to the supertransformations for
the fermions only. Complicated higher derivative corrections to the
supertransformations reappear when one tries to fix all local gauge
symmetries, breaking the gauge invariance. Note however that in such
construction the mass-like terms for the fermions in the Lagrangian
take a complicated non-diagonal form making calculations rather
cumbersome. Surprisingly however, in 4D the above results remain the
main results in the massive supersymmetric higher spin theory until
now\footnote{The attempts to developed the off-shell superfield
Lagrangian formulation of the massive higher spin supermultiplets
were realized only for some examples in \cite{BG1}, \cite{BG2},
\cite{BG3}, \cite{GKT}. General superfield formulation is still
undeveloped.}. The aim of this paper is to extend and generalize the
results of \cite{Zin07a} to the case of four dimensional $N=1$
$AdS_4$ superalgebra.

We use the gauge invariant description of the massive higher spin
bosonic and fermionic fields but in the frame-like version
\cite{Zin08b,PV10}. Recall that one of the attractive features of
such a formalism is that it  works nicely both in flat Minkowski
space as well as in $(A)dS$ spaces. Our strategy differs from that
of \cite{Zin07a}. For the Lagrangian we take just the sum of four
free Lagrangians for the two massive bosonic and two massive
fermionic fields entering the supermultiplet. Then for each pair of
bosonic and fermionic fields (we call it superblock in what follows)
we find the supertransformations leaving the sum of their two
Lagrangians invariant. Next we combine all four possible superblocks
and adjust their parameters so that the algebra of the
supertransformations is closed on-shell.

The paper is organized as follows. In section \ref{Section1} we give
all necessary descriptions of the frame-like formulation of massless
bosonic and fermionic higher spin fields and alos we present the
massless higher spin supermultiplets in $AdS_4$ in such a formalism.
Massless models given in this section will serve as the building
blocks for our construction of the massive higher spin models. In
section \ref{Section2} we give frame-like gauge invariant
formulations for free massive arbitrary integer and half-inter
spins. In section \ref{Section3} we consider massive superblocks
containing one massive bosonic and one massive fermionic field and
find corresponding supertransformations. In section \ref{Section4}
we combine the constructed massive superblocks into one massive
supermultiplet.

\noindent {\bf Notations and conventions.} In this work we use a
technique of $p$-forms taking the values in the Grassmann algebra.
The main geometrical objects are some $p$-forms $\Omega$
(p=0,1,2,3,4). They are defined as
$$
\Omega=dx^{\mu_1}\wedge...\wedge dx^{\mu_p}\Omega_{\mu_1...\mu_p},
$$
where $\Omega_{\mu_1...\mu_p}$ is the antisymmetric tensor field. In
particular, the partial derivative is defined as one-form
$d=dx^\mu\partial_\mu$. In 4D it is convenient to use a frame-like
multispinor formalism where all the Lorentz objects have local
totally symmetric dotted and undotted spinorial indices (see a
description of irreducible representations of 4D Lorentz group in
terms of dotted and undotted spinors e.g. in \cite{BK98}). To
simplify the expressions we will use the condensed notations for
them such that e.g.
$$
\Omega^{\alpha(m)\dot\alpha(n)} = \Omega^{(\alpha_1\alpha_2 \dots
\alpha_{m})(\dot\alpha_1\dot\alpha_2 \dots \dot\alpha_{n})}
$$
We also always assume that spinor indices denoted by the same
letters and placed on the same level are symmetrized, e.g.
$$
\Omega^{\alpha(m)\dot\alpha(n)} \zeta^{\dot\alpha} =
\Omega^{\alpha(m)(\dot\alpha_1\dots \dot\alpha_{n}}
\zeta^{\dot\alpha_{n+1})}
$$
We work in the $AdS_4$ space which is described by pair 1-forms:
background frame $e^{\alpha\dot\alpha}$ which enters explicitly in
all constructions and background Lorentz connections
$\omega^{\alpha\beta},\omega^{\dot\alpha\dot\beta}$ which are hidden
in the 1-form covariant derivative
$$
D\Omega^{\alpha(m)\dot\alpha(n)}=d\Omega^{\alpha(m)\dot\alpha(n)}
+\omega^\alpha{}_{\beta}\wedge\Omega^{\alpha(m-1)\beta\dot\alpha(n)}
+\omega^{\dot\alpha}{}_{\dot\beta}\wedge\Omega^{\alpha(m)\beta\dot\alpha(n-1)\dot\beta}.
$$
The covariant derivative satisfies the following normalization
conditions:
$$
D\wedge D \Omega^{\alpha(m)\dot\alpha(n)} = -
2\lambda^2(E^\alpha{}_\beta\wedge\Omega^{\alpha(m-1)\beta\dot\alpha(n)}+E^{\dot\alpha}{}_{\dot\beta}
\wedge\Omega^{\alpha(m)\dot\alpha(n-1)\dot\beta})
$$
where $E^{\alpha\beta},E^{\dot\alpha\dot\beta}$ are basis elements
of $2$-form spaces and defined below as the double product of
$e^{\alpha\dot\alpha}$.

Basis elements of $1,2,3,4$-form spaces are:
$$
e^{\alpha\dot\alpha},\quad E_2{}^{\alpha\beta},\quad
E_2{}^{\dot\alpha\dot\beta},\quad E_3{}^{\alpha\dot\alpha},\quad E_4
$$
They are defined as follows:
\begin{eqnarray*}
e^{\alpha\dot\alpha}\wedge  e^{\beta\dot\beta} &=&
\varepsilon^{\alpha\beta} E^{\dot\alpha\dot\beta} +
\varepsilon^{\dot\alpha\dot\beta }E^{\alpha\beta}
\\
E^{\alpha\beta}\wedge  e^{\gamma\dot\alpha} &=&
\varepsilon^{\alpha\gamma} E^{\beta\dot\alpha}+
\varepsilon^{\beta\gamma} E^{\alpha\dot\alpha}
\\
E^{\dot\alpha\dot\beta}\wedge  e^{\alpha\dot\gamma} &=&
-\varepsilon^{\dot\alpha\dot\gamma} E^{\alpha\dot\beta}
-\varepsilon^{\dot\beta\dot\gamma} E^{\alpha\dot\alpha}
\\
E^{\alpha\dot\alpha}\wedge  e^{\beta\dot\beta} &=&
\varepsilon^{\alpha\beta} \varepsilon^{\dot\alpha\dot\beta}E
\end{eqnarray*}
so the Hermitian conjugation laws look like
$$
(e^{\alpha\dot\alpha})^\dag=e^{\alpha\dot\alpha},\quad
(E^{\alpha(2)})^\dag=E^{\dot\alpha(2)},\quad
(E^{\alpha\dot\alpha})^\dag=-E^{\alpha\dot\alpha},\quad E^\dag=-E
$$
We also write some useful relations for these basis elements
$$
e^{\alpha}{}_{\dot\beta}\wedge  e^{\beta\dot\beta} = 2
E^{\alpha\beta},\qquad e_\beta{}^{\dot\alpha}\wedge
e^{\beta\dot\beta} = 2 E^{\dot\alpha\dot\beta}
$$
$$
E^{\alpha}{}_\gamma \wedge e^{\gamma\dot\alpha} = 3
E^{\alpha\dot\alpha},\qquad E^{\dot\alpha}{}_{\dot\gamma}\wedge
e^{\alpha\dot\gamma} = - 3E^{\alpha\dot\alpha}
$$
$$
E_\beta{}^{\dot\alpha}\wedge  e^{\beta\dot\beta} = 2
\varepsilon^{\dot\alpha\dot\beta}E, \qquad E^{\alpha}{}_{\dot\beta}
\wedge e^{\beta\dot\beta} = 2\varepsilon^{\alpha\beta} E
$$
$$
E^{\alpha\beta}\wedge  E^{\dot\alpha\dot\beta} = 0,\qquad
E^{\alpha}{}_\gamma \wedge E^{\beta\gamma} = 3
\varepsilon^{\alpha\beta} E
$$
The spinor indices are raised and lowered with the help of the
antisymmetric Lorentz invariant tensors $\epsilon^{\alpha\beta},
\epsilon^{\dot{\alpha}\dot{\beta}}$ and inverse
$\epsilon_{\alpha\beta}, \epsilon_{\dot{\alpha}\dot{\beta}}$
respectively. All the products  of $p$ - forms are understood in the
sense of wedge-products. Henceforth the sign of wedge product
$\wedge$ will be omitted.

\section{Massless higher spin models}\label{Section1}

In this section we provide all necessary description of the massless
bosonic and fermionic higher spin fields as well as massless higher
spin supermultiplets in the frame-like multispinor formalism used in
this work. In what follows they will serve as building blocks for
our construction for massive supermultiplets.

\subsection{Integer spin $k$}

In the frame-like formulation a massless spin-$k$ field ($k\geq2$)
is described by the physical one-form
$f^{\alpha(k-1)\dot\alpha(k-1)}$ and the auxiliary one-forms
$\Omega^{\alpha(k)\dot\alpha(k-2)},\Omega^{\alpha(k-2)\dot\alpha(k)}$,
being the higher spin generalization of the frame and Lorentz
connection in the frame-like formulation of gravity. Locally they
are two-component multispinors symmetric on their dotted and
undotted spinorial indices separately. These fields satisfy the
following reality condition
\begin{equation}\label{RealCond1}
(f^{\alpha(k-1)\dot\alpha(k-1)})^\dag =
f^{\alpha(k-1)\dot\alpha(k-1)},\qquad
(\Omega^{\alpha(k)\dot\alpha(k-2)})^\dag =
\Omega^{\alpha(k-2)\dot\alpha(k)}
\end{equation}
The free Lagrangian (a four-form in our formalism) for the massless
bosonic field in the four-dimensional $AdS_4$ space looks like this:
\begin{eqnarray}\label{MaslBosonLag}
\frac{(-1)^k}{i} {\cal L}_k &=& k
\Omega^{\alpha(k-1)\beta\dot\alpha(k-2)} E_\beta{}^\gamma
\Omega_{\alpha(k-1)\gamma\dot\alpha(k-2)} - (k-2)
\Omega^{\alpha(k)\dot\alpha(k-3)\dot\beta}
E_{\dot\beta}{}^{\dot\gamma}
\Omega_{\alpha(k)\dot\alpha(k-3)\dot\gamma} \nonumber
\\
 && + 2 \Omega^{\alpha(k-1)\beta\dot\alpha(k-2)} e_\beta{}^{\dot\beta}
Df_{\alpha(k-1)\dot\alpha(k-2)\dot\beta} \nonumber
\\
 && + 2k\lambda^2 f^{\alpha(k-2)\beta\dot\alpha(k-1)} E_\beta{}^\gamma
f_{\alpha(k-2)\gamma\dot\alpha(k-1)} + h.c.
\end{eqnarray}
Here and henceforth $h.c.$ means hermitian conjugate terms defined
by rules (\ref{RealCond1}). This Lagrangian is invariant under the
following gauge transformations
\begin{eqnarray}\label{MaslGTB}
\delta \Omega^{\alpha(k)\dot\alpha(k-2)} &=& D
\eta^{\alpha(k),\dot\alpha(k-2)} +
e_\beta{}^{\dot\alpha}\zeta^{\alpha(k)\beta\dot\alpha(k-3)} +
\lambda e^\alpha{}_{\dot\beta}
\xi^{\alpha(k-1)\dot\alpha(k-2)\dot\beta}\nonumber
\\
\delta \Omega^{\alpha(k-2)\dot\alpha(k)} &=& D
\eta^{\alpha(k-2),\dot\alpha(k)} +
e^\alpha{}_{\dot\beta}\zeta^{\alpha(k-3)\dot\alpha(k)\dot\beta} +
\lambda e_\beta{}^{\dot\alpha} \xi^{\alpha(k-2)\beta\dot\alpha(k-1)}
\nonumber
\\
\delta f^{\alpha(k-1)\dot\alpha(k-1)} &=&
D\xi^{\alpha(k-1)\dot\alpha(k-1)} +
e_\beta{}^{\dot\alpha}\eta^{\alpha(k-1)\beta\dot\alpha(k-2)} +
e^\alpha{}_{\dot\beta}\eta^{\alpha(k-2)\dot\alpha(k-1)\dot\beta}
\end{eqnarray}
where zero-forms $\xi^{\alpha(k-1)\dot\alpha(k-1)}$ and
$\eta^{\alpha(k),\dot\alpha(k-2)}$ are the gauge parameters for the
gauge fields $f^{\alpha(k-1)\dot\alpha(k-1)}$ and
$\Omega^{\alpha(k),\dot\alpha(k-2)}$. The additional gauge parameter
$\zeta^{\alpha(k+1)\dot\alpha(k-3)}$ leads to the introduction of
the so-called extra field $\Upsilon^{\alpha(k+1)\dot\alpha(k-3)} +
h.c.$ which in turn requires introduction of next extra symmetries
and so on. The procedure stops at
\begin{eqnarray}\label{ExtMaslGTB}
\Upsilon^{\alpha(k+t)\dot\alpha(k-t-2)},\quad
\Upsilon^{\alpha(k-t-2)\dot\alpha(k+t)},\qquad 1\leq t\leq k-2
\end{eqnarray}
These extra gauge fields do not enter the free Lagrangian but play
an important role in non-linear higher spin theory.

One of the nice features of the frame-like formulation is that for
all fields (physical, auxiliary and extra ones) one can construct a
gauge invariant two-form (curvature) generalizing curvature and
torsion for gravity. For the physical and auxiliary fields they have
the form
\begin{eqnarray}\label{MaslBosonCurv}
{\cal R}^{\alpha(k)\dot\alpha(k-2)} &=& D
\Omega^{\alpha(k),\dot\alpha(k-2)} + \lambda e^\alpha{}_{\dot\beta}
f^{\alpha(k-1)\dot\alpha(k-2)\dot\beta} + e_\beta{}^{\dot\alpha}
\Upsilon^{\alpha(k)\beta\dot\alpha(k-3)}\nonumber
\\
{\cal R}^{\alpha(k-2)\dot\alpha(k)} &=& D
\Omega^{\alpha(k-2),\dot\alpha(k)} + \lambda
e_\beta{}^{\dot\alpha}f^{\alpha(k-2)\beta\dot\alpha(k-1)} +
e^\alpha{}_{\dot\beta}\Upsilon^{\alpha(k-3)\dot\alpha(k)\dot\beta}
\nonumber
\\
{\cal T}^{\alpha(k-1)\dot\alpha(k-1)} &=& D
f^{\alpha(k-1)\dot\alpha(k-1)} +
e_\beta{}^{\dot\alpha}\Omega^{\alpha(k-1)\beta\dot\alpha(k-2)} +
e^\alpha{}_{\dot\beta}\Omega^{\alpha(k-2)\dot\alpha(k-1)\dot\beta}
\end{eqnarray}
In our construction for the massless and massive supermultiplets we
use only the physical and auxiliary fields working in the so-called
1 and 1/2 order formalism which is very well known in supergravity.
Namely, we do not consider any variations of the auxiliary fields
but all calculations are done using the "zero torsion condition":
\begin{eqnarray}\label{MaslOn-Shell}
{\cal T}^{\alpha(k-1)\dot\alpha(k-1)} \approx 0
\quad\Rightarrow\quad e_\beta{}^{\dot\alpha} {\cal
R}^{\alpha(k-1)\beta\dot\alpha(k-2)} + e^\alpha{}_{\dot\beta}{\cal
R}^{\alpha(k-2)\dot\alpha(k-1)\dot\beta}\approx 0
\end{eqnarray}
At the same time the variation of the Lagrangian
(\ref{MaslBosonLag}) under the arbitrary variations of the physical
fields can be written in the following simple form
\begin{eqnarray}\label{MaslBosonVar}
(-1)^k\delta{\cal L}_k&=&-i2{\cal
R}^{\alpha(k-1)\beta\dot\alpha(k-2)}e_\beta{}^{\dot\beta} \delta
f_{\alpha(k-1)\dot\alpha(k-2)\dot\beta}+h.c.
\end{eqnarray}
One can see that in (\ref{MaslOn-Shell}) and (\ref{MaslBosonVar}),
the curvatures ${\cal R}$ enter in such combinations that extra
gauge field $\Upsilon$ is dropped out, therefore below they will be
omitted.

\subsection{Half-integer spin $k+1/2$}

In the frame-like formulation, the massless spin-$(k+1/2)$ field
($k\geq1$) is described by physical 1-forms
$\Phi^{\alpha(k)\dot\alpha(k-1)},\Phi^{\alpha(k-1)\dot\alpha(k)}$.
As in the bosonic case, these two-component multispinors are
symmetric on their dotted and undotted spinorial indices separately
and satisfy the reality condition
$$
(\Phi^{\alpha(k)\dot\alpha(k-1)})^\dag =
\Phi^{\alpha(k-1)\dot\alpha(k)}
$$
The free Lagrangian for such fields in $AdS_4$ space looks like
this:
\begin{eqnarray}\label{MaslFermLag}
(-1)^k{\cal L}_{k+\frac12} &=&
\Psi_{\alpha(k-1)\beta\dot\alpha(k-1)}e^\beta{}_{\dot\beta}
D\Psi^{\alpha(k-1)\dot\alpha(k-1)\dot\beta}\nonumber
\\
 && + d_{k-1}[(k+1) \Psi_{\alpha(k-1)\beta\dot\alpha(k-1)}
E^\beta{}_{\gamma} \Psi^{\alpha(k-1)\gamma\dot\alpha(k-1)} \nonumber
\\
 && - (k-1)\Psi_{\alpha(k)\dot\alpha(k-2)\dot\beta}
E^{\dot\beta}{}_{\dot\gamma}
\Psi^{\alpha(k)\dot\alpha(k-2)\dot\gamma}+ h.c.]
\end{eqnarray}
and is invariant under the following gauge transformation
\begin{eqnarray}\label{MaslFermGT}
\delta \Psi^{\alpha(k)\dot\alpha(k-1)} &=& D
\xi^{\alpha(k)\dot\alpha(k-1)} + e_\beta{}^{\dot\alpha}
\eta^{\alpha(k)\beta\dot\alpha(k-2)} + 2d_{k-1}
e^\alpha{}_{\dot\beta}\xi^{\alpha(k-1)\dot\alpha(k-1)\dot\beta}
\nonumber
\\
\delta \Psi^{\alpha(k-1)\dot\alpha(k)} &=& D
\xi^{\alpha(k-1)\dot\alpha(k)} +
e^\alpha{}_{\dot\beta}\eta^{\alpha(k-2)\dot\alpha(k)\dot\beta} +
2d_{k-1} e_\beta{}^{\dot\alpha}
\xi^{\alpha(k-1)\beta\dot\alpha(k-1)}
\end{eqnarray}
where
$$
d_{k-1} = \pm \frac{\lambda}{2}
$$
Note that the transformations with the gauge parameters
$\eta^{\alpha(k+1)\dot\alpha(k-2)},\eta^{\alpha(k-2)\dot\alpha(k+1)}$
lead to the introduction of extra fields that play the same role as
in the bosonic case and do not enter the free Lagrangian. Up to
these extra fields the gauge invariant curvatures for the physical
fermionic fields have the following form
\begin{eqnarray}\label{MaslFermCurv}
{\cal F}^{\alpha(k)\dot\alpha(k-1)} &=& D
\Psi^{\alpha(k)\dot\alpha(k-1)} + 2d_{k-1}
e^\alpha{}_{\dot\beta}\Psi^{\alpha(k-1)\dot\alpha(k-1)\dot\beta}
\nonumber
\\
{\cal F}^{\alpha(k-1)\dot\alpha(k)} &=& D
\Psi^{\alpha(k-1)\dot\alpha(k)} + 2d_{k-1}
e_\beta{}^{\dot\alpha}\Psi^{\alpha(k-1)\beta\dot\alpha(k-1)}
\end{eqnarray}
The variation of the free Lagrangian (\ref{MaslFermLag}) under the
arbitrary variations of the physical fields can be written as
\begin{eqnarray}\label{MaslFermVar}
(-1)^k\delta{\cal L}_{k+\frac12}&=&-{\cal
F}_{\alpha(k-1)\beta\dot\alpha(k-1)}e^\beta{}_{\dot\beta}
\delta\Psi^{\alpha(k-1)\dot\alpha(k-1)\dot\beta}+h.c.
\end{eqnarray}
Here the curvature ${\cal F}$ also enters in such a combination with
the frame $e$ that the extra fields drop out.

In the following part of this section we combine massless higher
spin fermionic and bosonic fields in the $N=1$ supermultiplet and
construct an explicit form of the supertransformations leaving the
sum of the free Lagrangians invariant.

\subsection{Supermultiplet $(k+1/2,k)$}

This supermultiplet contains higher half-integer spin $k+1/2$ and
integer spin $k$. They are described by
$(\Phi^{\alpha(k)\dot\alpha(k-1)}, h.c.)$ and
$(f^{\alpha(k-1)\dot\alpha(k-1)},
\Omega^{\alpha(k)\dot\alpha(k-2)},h.c.)$ respectively. We choose an
ansatz for the supertransformations in the following form (as it was
already mentioned, we consider supertransformations for the physical
fields only):
\begin{eqnarray}\label{MaslST1}
\delta f^{\alpha(k-1)\dot\alpha(k-1)} &=&
\alpha_{k-1}\Phi^{\alpha(k-1)\beta\dot\alpha(k-1)}\zeta_\beta -
\bar\alpha_{k-1}\Phi^{\alpha(k-1)\dot\alpha(k-1)\dot\beta}
\zeta_{\dot\beta} \nonumber
\\
\delta \Phi^{\alpha(k)\dot\alpha(k-1)} &=&
\beta_{k-1}\Omega^{\alpha(k)\dot\alpha(k-2)} \zeta^{\dot\alpha} +
\gamma_{k-1} f^{\alpha(k-1)\dot\alpha(k-1)} \zeta^{\alpha} \nonumber
\\
\delta \Phi^{\alpha(k-1)\dot\alpha(k)} &=&
\bar\beta_{k-1}\Omega^{\alpha(k-2)\dot\alpha(k)}\zeta^{\alpha} +
\bar\gamma_{k-1}f^{\alpha(k-1)\dot\alpha(k-1)} \zeta^{\dot\alpha}
\end{eqnarray}
were we assume that the coefficients $\alpha_k,\beta_k,\gamma_k$ are
complex. The parameters of the supertransformation $\zeta^\alpha,
\zeta^{\dot\alpha}$ in $AdS_4$ satisfy the relations
\begin{equation}\label{ParamRealat}
D \zeta^\alpha = - \lambda e^\alpha{}_{\dot\beta} \zeta^{\dot\beta},
\qquad D \zeta^{\dot\alpha} = - \lambda
e_\beta{}^{\dot\alpha}\zeta^{\beta}
\end{equation}
Using the expressions for Lagrangian variations (\ref{MaslBosonVar})
and (\ref{MaslFermVar}) as well as on-shell identity
(\ref{MaslOn-Shell}) the variation for the sum of the bosonic and
fermionic Lagrangians can be written as follows:
\begin{eqnarray}\label{MaslVar1}
(-1)^k\delta({\cal L}_k+{\cal
L}_{k+\frac12})&=&4i\alpha_{k-1}\Phi_{\alpha(k-2)\beta\gamma\dot\alpha(k-1)}e^\gamma{}_{\dot\gamma}
{\cal R}^{\alpha(k-2)\dot\alpha(k-1)\dot\gamma}\zeta^{\beta}\nn\\
&&- (k-1)\bar\beta_{k-1}{\cal
F}_{\alpha(k-1)\beta\dot\alpha(k-1)}e^\beta{}_{\dot\beta}
\Omega^{\alpha(k-2)\dot\alpha(k-1)\dot\beta}\zeta^{\alpha}\nn
\\
&& -\bar\gamma_{k-1}({\cal
F}_{\alpha(k-1)\beta\dot\alpha(k-1)}e^\beta{}_{\dot\beta}
f^{\alpha(k-1)\dot\alpha(k-1)}\zeta^{\dot\beta}\nn\\
&&+ (k-1){\cal
F}_{\alpha(k-1)\beta\dot\alpha(k-1)}e^\beta{}_{\dot\beta}
f^{\alpha(k-1)\dot\alpha(k-2)\dot\beta}\zeta^{\dot\alpha})+h.c.
\end{eqnarray}
Note that the invariance of the Lagrangian under the
supertransformations can be achieved up to the total derivative only
and this leads to a number of useful identities. For example, let us
consider
\begin{eqnarray*}
0 &\approx& D[\Phi_{\alpha(k-1)\beta\dot\alpha(k-1)}
e^\beta{}_{\dot\beta}
\Omega^{\alpha(k-2)\dot\alpha(k-1)\dot\beta} \zeta^{\alpha}] \\
 &=& D \Phi_{\alpha(k-1)\beta\dot\alpha(k-1)}
e^\beta{}_{\dot\beta} \Omega^{\alpha(k-2)\dot\alpha(k-1)\dot\beta}
\zeta^{\alpha} + \Phi_{\alpha(k-1)\beta\dot\alpha(k-1)}
e^\beta{}_{\dot\beta} D
\Omega^{\alpha(k-2)\dot\alpha(k-1)\dot\beta} \zeta^{\alpha} \\
 && - \Phi_{\alpha(k-1)\beta\dot\alpha(k-1)}
e^\beta{}_{\dot\beta} \Omega^{\alpha(k-2)\dot\alpha(k-1)\dot\beta} D
\zeta^{\alpha}
\end{eqnarray*}
Using the explicit expressions for the bosonic (\ref{MaslBosonCurv})
and fermionic (\ref{MaslFermCurv}) curvatures as well as relation
(\ref{ParamRealat}) we obtain:
\begin{eqnarray*}
0 &=& {\cal F}_{\alpha(k-1)\beta\dot\alpha(k-1)}
e^\beta{}_{\dot\beta} \Omega^{\alpha(k-2)\dot\alpha(k-1)\dot\beta}
\zeta^{\alpha} + \Phi_{\alpha(k-2)\beta\gamma\dot\alpha(k-1)}
e^\beta{}_{\dot\beta} {\cal R}^{\alpha(k-2)\dot\alpha(k-1)\dot\beta}
\zeta^{\gamma}
\\
 && - 2{b_{k-1}} [(k+1)E^{\dot\beta}{}_{\dot\gamma}
\Psi^{\alpha(k-2)\gamma\dot\alpha(k-1)\dot\gamma}
\Omega_{\alpha(k-2)\dot\alpha(k-1)\dot\beta} \zeta_{\gamma}
\\
 && - (k-2) E_\beta{}^\alpha
\Psi^{\alpha(k-3)\beta\gamma\dot\alpha(k)}
\Omega_{\alpha(k-2)\dot\alpha(k)}\zeta_{\gamma} - E_\beta{}^\gamma
\Psi^{\alpha(k-2)\beta\dot\alpha(k)}
\Omega_{\alpha(k-2)\dot\alpha(k)} \zeta_{\gamma} ]
\\
&& - \lambda^2 [(k+1)E^\beta{}_\delta
\Phi_{\alpha(k-2)\beta\gamma\dot\alpha(k-1)}
f^{\alpha(k-2)\delta\dot\alpha(k-1)} - (k-1)
E_{\dot\beta}{}^{\dot\alpha}
\Phi_{\alpha(k-1)\gamma\dot\alpha(k-1)}f^{\alpha(k-1)\dot\alpha(k-2)\dot\beta}
]\zeta^{\gamma}
\\
 && + \lambda [E^{\alpha\beta} \Phi_{\alpha(k-1)\beta\dot\alpha(k-1)}
\Omega^{\alpha(k-2)\dot\alpha(k-1)\dot\beta} \zeta_{\dot\beta}]
\end{eqnarray*}
Similarly, if one considers two relations:
\begin{eqnarray*}
0 &\approx& D[\Phi_{\alpha(k-1)\beta\dot\alpha(k-1)}
e^\beta{}_{\dot\beta} f^{\alpha(k-1)\dot\alpha(k-1)}
\zeta^{\dot\beta}]
\\
0 &\approx& D[\Phi_{\alpha(k-1)\beta\dot\alpha(k-2)\dot\gamma}
e^\beta{}_{\dot\beta}
f^{\alpha(k-1)\dot\alpha(k-2)\dot\beta}\zeta^{\dot\gamma}]
\end{eqnarray*}
then using the explicit expression for the fermionic curvature as
well as zero torsion condition one obtains:
\begin{eqnarray*}
0 &=& {\cal F}_{\alpha(k-1)\beta\dot\alpha(k-1)}
e^\beta{}_{\dot\beta}
f^{\alpha(k-1)\dot\alpha(k-1)}\zeta^{\dot\beta}
\\
 && - 2{b_{k-1}} [(k+1)E^{\dot\beta}{}_{\dot\gamma}
\Psi^{\alpha(k-1)\dot\alpha(k-1)\dot\gamma}
f_{\alpha(k-1)\dot\alpha(k-1)} \zeta_{\dot\beta}
\\
 && - (k-1)E_\beta{}^\alpha
\Psi^{\alpha(k-2)\beta\dot\alpha(k-1)\dot\beta}
f_{\alpha(k-1)\dot\alpha(k-1)} \zeta_{\dot\beta} ]
\\
 && - [ (k-1)(- E_{\dot\beta}{}^{\dot\alpha}
\Phi_{\alpha(k)\dot\alpha(k-1)} \Omega^{\alpha(k)\dot\alpha(k-2)} +
E^\beta{}_\gamma \Phi_{\alpha(k-1)\beta\dot\alpha(k-2)\dot\beta}
\Omega^{\alpha(k-1)\gamma\dot\alpha(k-2)}) \zeta^{\dot\beta}
\\
 && + (k-1)E^{\beta\alpha}
\Phi_{\alpha(k-1)\beta\dot\alpha(k-1)}
\Omega^{\alpha(k-2)\dot\alpha(k-1)\dot\gamma} \zeta_{\dot\gamma}]
\\
 && - 2\lambda E_\gamma{}^\beta
\Phi_{\alpha(k-1)\beta\dot\alpha(k-1)}
f^{\alpha(k-1)\dot\alpha(k-1)} \zeta^{\gamma}
\end{eqnarray*}
\begin{eqnarray*}
0 &=& {\cal F}_{\alpha(k-1)\beta\dot\alpha(k-2)\dot\gamma}
e^\beta{}_{\dot\beta}
f^{\alpha(k-1)\dot\alpha(k-2)\dot\beta}\zeta^{\dot\gamma}
\\
 && - 2{b_{k-1}} [(k+1)E^{\dot\beta}{}_{\dot\delta}
\Psi^{\alpha(k-1)\dot\alpha(k-2)\dot\gamma\dot\delta}
f_{\alpha(k-1)\dot\alpha(k-2)\dot\beta} \zeta_{\dot\gamma}
\\
 && - (k-1)E_\beta{}^\alpha
\Psi^{\alpha(k-2)\beta\dot\alpha(k-1)\dot\gamma}
f_{\alpha(k-1)\dot\alpha(k-1)} \zeta_{\dot\gamma}]
\\
 && - [kE^\beta{}_\delta
\Phi_{\alpha(k-1)\beta\dot\alpha(k-2)\dot\gamma}
\Omega^{\alpha(k-1)\delta\dot\alpha(k-2)} -(k-2)
E_{\dot\beta}{}^{\dot\alpha}
\Phi_{\alpha(k)\dot\alpha(k-2)\dot\gamma}
\Omega^{\alpha(k)\dot\alpha(k-3)\dot\beta} ]\zeta^{\dot\gamma}
\\
 && + \lambda[ E^{\dot\gamma}{}_{\dot\beta}
\Phi_{\alpha(k-1)\gamma\dot\alpha(k-2)\dot\gamma}
f^{\alpha(k-1)\dot\alpha(k-2)\dot\beta} \zeta^{\gamma} -
E_\gamma{}^\beta \Phi_{\alpha(k-1)\beta\dot\alpha(k-1)}
f^{\alpha(k-1)\dot\alpha(k-1)} \zeta^{\gamma}]
\end{eqnarray*}
Using these identities we obtain from the requirement $\delta({\cal
L}_k+{\cal L}_{k+\frac12}) = 0$:
\begin{equation}\label{MaslSol1}
\alpha_{k-1} = i\frac{(k-1)}{4} \bar\beta_{k-1}, \qquad \gamma_{k-1}
= \lambda \bar\beta_{k-1}, \qquad 2d_{k-1}\bar\beta_{k-1} = \lambda
\beta_{k-1}
\end{equation}
The solution of the last relation depends on the sign of
$d_{k-1}=\pm\frac{\lambda}{2}$. The parameter $\beta_{k}$ is real
for the "+" sign and is imaginary for the  "-". These two solutions
correspond to the parity-even and parity-odd bosonic fields entering
the supermultiplet. This fact will be important for the construction
of massive supermultiplets where two bosonic fields must have
opposite parities.

\subsection{Supermultiplet $(k,k-1/2)$}

This supermultiplet contains higher integer spin $k$ and
half-integer spin $k-1/2$. They are described by
$(f^{\alpha(k-1)\dot\alpha(k-1)},\Omega^{\alpha(k)\dot\alpha(k-2)},h.c.)$
and $(\Phi^{\alpha(k-1)\dot\alpha(k-2)}, h.c.)$ respectively. We
choose an ansatz for the supertransformations in the following form
\begin{eqnarray}\label{MaslST2}
\delta f^{\alpha(k-1)\dot\alpha(k-1)} &=&
\alpha'_{k-1}\Phi^{\alpha(k-1)\dot\alpha(k-2)} \zeta^{\dot\alpha} -
\bar\alpha'_{k-1} \Phi^{\alpha(k-2)\dot\alpha(k-1)} \zeta^{\alpha}
\nonumber
\\
\delta \Psi^{\alpha(k-1)\dot\alpha(k-2)} &=&
\beta'_{k-1}\Omega^{\alpha(k-1)\beta\dot\alpha(k-2)} \zeta_\beta +
\gamma'_{k-1}f^{\alpha(k-1)\dot\alpha(k-2)\dot\beta}
\zeta_{\dot\beta} \nonumber
\\
\delta \Psi^{\alpha(k-2)\dot\alpha(k-1)} &=&
\bar\beta'_{k-1}\Omega^{\alpha(k-2)\dot\alpha(k-1)\dot\beta}
\zeta_{\dot\beta} + \bar\gamma'_{k-1}
f^{\alpha(k-2)\beta\dot\alpha(k-1)} \zeta_\beta
\end{eqnarray}
Here in the most general case, coefficients
$\alpha'_k,\beta'_k,\gamma'_k$ are complex. Using the expressions
for Lagrangian variations (\ref{MaslBosonVar}) and
(\ref{MaslFermVar}) as well as on-shell relation
(\ref{MaslOn-Shell})  we get
\begin{eqnarray*}
(-1)^k \delta ({\cal L}_k+{\cal L}_{k-\frac12}) &=& -
4i(k-1)\alpha'_{k-1} \Phi_{\alpha(k-2)\beta\dot\alpha(k-2)}
e^\beta{}_{\dot\beta} {\cal
R}^{\alpha(k-2)\dot\alpha(k-2)\dot\beta\dot\gamma}
\zeta_{\dot\gamma}
\\
 && + \beta'^*_{k-1}{\cal F}_{\alpha(k-2)\gamma\dot\alpha(k-2)}
e^\gamma{}_{\dot\gamma}
\Omega^{\alpha(k-2)\dot\alpha(k-2)\dot\gamma\dot\beta}
\zeta_{\dot\beta}
\\
 && + \gamma'^*_{k-1}{\cal F}_{\alpha(k-2)\gamma\dot\alpha(k-2)}
e^\gamma{}_{\dot\gamma}
f^{\alpha(k-2)\beta\dot\alpha(k-2)\dot\gamma} \zeta_\beta + h.c.
\end{eqnarray*}
As in previous case from two relations
\begin{eqnarray*}
0 &\approx& D [\Phi_{\alpha(k-2)\gamma\dot\alpha(k-2)}
e^\gamma{}_{\dot\gamma}
\Omega^{\alpha(k-2)\dot\alpha(k-2)\dot\gamma\dot\beta}
\zeta_{\dot\beta}]
\\
0 &\approx& D [\Phi_{\alpha(k-2)\gamma\dot\alpha(k-2)}
e^\gamma{}_{\dot\gamma}
f^{\alpha(k-2)\beta\dot\alpha(k-2)\dot\gamma} \zeta_\beta]
\end{eqnarray*}
one can derive two identities:
\begin{eqnarray*}
0 &=& {\cal F}_{\alpha(k-2)\gamma\dot\alpha(k-2)}
e^\gamma{}_{\dot\gamma}
\Omega^{\alpha(k-2)\dot\alpha(k-2)\dot\gamma\dot\beta}
\zeta_{\dot\beta} + \Phi_{\alpha(k-2)\gamma\dot\alpha(k-2)}
e^\gamma{}_{\dot\gamma} {\cal
R}^{\alpha(k-2)\dot\alpha(k-2)\dot\gamma\dot\beta} \zeta_{\dot\beta}
\\
 && + 2{b_{k-2}}[k E^{\dot\gamma}{}_{\dot\delta}
\Psi^{\alpha(k-2)\dot\alpha(k-2)\dot\delta}
\Omega_{\alpha(k-2)\dot\alpha(k-2)\dot\gamma\dot\beta}
\zeta^{\dot\beta}
\\
 && - (k-2)E_\gamma{}^\beta \Psi^{\alpha(k-3)\gamma\dot\alpha(k-1)}
\Omega_{\alpha(k-3)\beta\dot\alpha(k-1)\dot\beta} \zeta^{\dot\beta}]
\\
 && - \lambda^2 [- (k-2)E_{\dot\gamma}{}^{\dot\alpha}
\Phi_{\alpha(k-1)\dot\alpha(k-2)}
f^{\alpha(k-1)\dot\alpha(k-3)\dot\beta\dot\gamma} + (k+1)
E^\gamma{}_\beta \Phi_{\alpha(k-2)\gamma\dot\alpha(k-2)}
f^{\alpha(k-2)\beta\dot\alpha(k-2)\dot\beta})
\\
 && - E_{\dot\gamma}{}^{\dot\beta} \Phi_{\alpha(k-1)\dot\alpha(k-2)}
f^{\alpha(k-1)\dot\alpha(k-2)\dot\gamma}] \zeta_{\dot\beta}
\\
 && + \lambda [E_{\dot\beta\dot\gamma}
\Phi_{\alpha(k-2)\gamma\dot\alpha(k-2)}
\Omega^{\alpha(k-2)\dot\alpha(k-2)\dot\gamma\dot\beta} \zeta^\gamma]
\end{eqnarray*}
\begin{eqnarray*}
0 &=& {\cal F}_{\alpha(k-2)\gamma\dot\alpha(k-2)}
e^\gamma{}_{\dot\gamma}
f^{\alpha(k-2)\beta\dot\alpha(k-2)\dot\gamma}\zeta_\beta
\\
 && + 2{b_{k-2}}[k E^{\dot\gamma}{}_{\dot\beta}
\Psi^{\alpha(k-2)\dot\alpha(k-2)\dot\beta}
f_{\alpha(k-2)\beta\dot\alpha(k-2)\dot\gamma} \zeta^\beta
\\
 && - (k-2)E_\gamma{}^\alpha \Psi^{\alpha(k-3)\gamma\dot\alpha(k-1)}
f_{\alpha(k-2)\beta\dot\alpha(k-1)} \zeta^\beta]
\\
 && - [-(k-2) E_{\dot\gamma}{}^{\dot\alpha}
\Phi_{\alpha(k-1)\dot\alpha(k-2)}
\Omega^{\alpha(k-1)\beta\dot\alpha(k-3)\dot\gamma} \zeta_\beta
\\
 && + kE^\gamma{}_\delta
\Phi_{\alpha(k-2)\gamma\dot\alpha(k-2)}\Omega^{\alpha(k-2)\beta\delta\dot\alpha(k-2)}
\zeta_\beta + E_{\dot\gamma\dot\beta}
\Phi_{\alpha(k-2)\gamma\dot\alpha(k-2)}
\Omega^{\alpha(k-2)\dot\alpha(k-2)\dot\gamma\dot\beta} \zeta^\gamma]
\\
 && + \lambda[ E_{\dot\beta\dot\gamma}
\Phi_{\alpha(k-1)\dot\alpha(k-2)}
f^{\alpha(k-1)\dot\alpha(k-2)\dot\gamma} \zeta^{\dot\beta} +
E_\beta{}^\gamma \Phi_{\alpha(k-2)\gamma\dot\alpha(k-2)}
f^{\alpha(k-2)\beta\dot\alpha(k-2)\dot\gamma} \zeta_{\dot\gamma} ]
\end{eqnarray*}
Then the invariance of the Lagrangian under the supertransformations
requires that
\begin{equation}\label{MaslSol2}
\alpha'_{k-1} = \frac{i}{4(k-1)} \bar\beta'_{k-1}, \qquad
\gamma'_{k-1} = \lambda \bar\beta'_{k-1}, \qquad 2d_{k-1}
\bar\beta'_{k-1} = \lambda \beta'_{k-1}
\end{equation}
As in the previous case, we see from last relation that parameter
$\beta'_{k}$ can be real or imaginary. It depends on the sign of
$d_{k-1}=\pm\frac{\lambda}{2}$ and is related to the parity of the
fields entering the supermultiplet.

\section{Massive higher spin fields}\label{Section2}

In this section we provide frame-like gauge invariant formulation
for massive arbitrary integer and half-inter spins \cite{Zin08b} but
with the multispinor formalism used for all local indices.

\subsection{Integer spin $s$}\label{MasbBoson}

In the gauge invariant formalism a massive integer spin-$s$ field is
described by a set of massless fields with spins $0\leq k\leq s$.
Frame-like formulation of massless bosonic fields with spins
$k\geq2$ were considered above, they are described by one-forms
$(f^{\alpha(k-1)\dot\alpha(k-1)},\Omega^{\alpha(k)\dot\alpha(k-2)} +
h.c.)$ while massless spin-$1$ is described by the physical one-form
$A$ and auxiliary zero-forms $B^{\alpha(2)},B^{\dot\alpha(2)}$, and
massless spin-$0$ is described by physical zero-form $\varphi$ and
auxiliary zero-form $\pi^{\alpha\dot\alpha}$.

The gauge invariant Lagrangian for the massive bosonic field has the
form:
\begin{equation}\label{MasvBosonLag}
{\cal L} = {\cal L}_{kin} + {\cal L}_{cross} + {\cal L}_{mass}
\end{equation}
\begin{eqnarray*}
\frac{1}{i}{\cal L}_{kin} &=& \sum_{k=2}^{s}(-1)^k {\cal L}_k + 4
EB_{\alpha(2)} B^{\alpha(2)} +2 E_{\alpha(2)}B^{\alpha(2)} DA + h.c.
\\
 && - 6E \pi_{\alpha\dot\alpha} \pi^{\alpha\dot\alpha}
-12 E_{\alpha\dot\alpha} \pi^{\alpha\dot\alpha} D\varphi \\
\frac{1}{i}{\cal L}_{cross} &=&
\sum_{k=3}^{s}(-1)^{k+1}a_k[E_{\beta(2)}
\Omega^{\alpha(k-2)\beta(2)\dot\alpha(k-2)}
f_{\alpha(k-2)\dot\alpha(k-2)}
\\
 && + \frac{(k-2)}{k}E_{\beta(2)}
f^{\alpha(k-3)\beta(2)\dot\alpha(k-1)}
\Omega_{\alpha(k-3)\dot\alpha(k-1)} + h.c.]
\\
 && + a_0[\Omega^{\alpha(2)} E_{\alpha(2)} A - 2 B^{\beta\alpha}
E_\beta{}^{\dot\beta} f_{\alpha\dot\beta} + h.c.] +
\tilde{a}_0 E_{\alpha\dot\alpha} \pi^{\alpha\dot\alpha} A \\
\frac{1}{i}{\cal L}_{mass} &=& \sum_{k=2}^s(-1)^kb_k[
f^{\alpha(k-2)\beta\dot\alpha(k-1)} E_\beta{}^\gamma
f_{\alpha(k-2)\gamma\dot\alpha(k-1)} + h.c.]
\\
 && + \frac{a_0\tilde{a}_0}{2} E_{\alpha\dot\alpha}
f^{\alpha\dot\alpha}\varphi + 3a_0{}^2 E \varphi^2
\end{eqnarray*}
where
\begin{eqnarray}
b_k &=& \frac{2s(s+1)}{k(k-1)(k+1)}M^2, \qquad M{}^2 =
m{}^2+s(s-1)\lambda^2 \nonumber
\\
a_k{}^2 &=& \frac{4(s-k+1)(s+k)}{(k-1)(k-2)} [M^2-k(k-1)\lambda^2]
\label{boson_date}
\\
a_0{}^2 &=& 2(s-1)(s+2) [M^2-2\lambda^2], \qquad \tilde{a}_0{}^2 =
24s(s+1)M^2 \nonumber
\end{eqnarray}
Here ${\cal L}_{kinetic}$ is just the sum of kinetic terms for all
fields, that for $k \ge 2$ were defined in (\ref{MaslBosonLag}),
${\cal L}_{mas}$ is the sum of the mass terms for them, while ${\cal
L}_{cross}$ contains cross-terms gluing all these fields together.
In what follows we assume that all parameters $a_k,a_0,\tilde{a}_0$
are positive.

Explicit form of the coefficients (\ref{boson_date}) are determined
by the invariance of the Lagrangian (\ref{MasvBosonLag}) under the
following gauge transformations
\begin{eqnarray}\label{MasvBosonGT}
\delta f^{\alpha(k-1)\dot\alpha(k-1)} &=& D
\xi^{\alpha(k-1)\dot\alpha(k-1)} +
e_\beta{}^{\dot\alpha}\eta^{\alpha(k-1)\beta\dot\alpha(k-2)} +
e^\alpha{}_{\dot\beta}\eta^{\alpha(k-2)\dot\alpha(k-1)\dot\beta}
\nonumber
\\
 && + \frac{(k-1)a_{k+1}}{2(k+1)} e_{\beta\dot\beta}
\xi^{\alpha(k-1)\beta\dot\alpha(k-1)\dot\beta} + \frac{a_k}{2k(k-1)}
e^{\alpha\dot\alpha} \xi^{\alpha(k-2)\dot\alpha(k-2)} \nonumber
\\
\delta \Omega^{\alpha(k),\dot\alpha(k-2)} &=& D
\eta^{\alpha(k),\dot\alpha(k-2)} + \frac{a_{k+1}}{2}
e_{\beta\dot\beta} \eta^{\alpha(k)\beta\dot\alpha(k-2)\dot\beta} +
\frac{a_k}{2k(k+1)} e^{\alpha\dot\alpha}
\eta^{\alpha(k-1)\dot\alpha(k-3)} \nonumber
\\
 && + \frac{b_k}{2k} e^\alpha{}_{\dot\beta}
\xi^{\alpha(k-1)\dot\alpha(k-2)\dot\beta}
\\
\delta f^{\alpha\dot\alpha} &=& D\xi^{\alpha\dot\alpha} +
e_\beta{}^{\dot\alpha} \eta^{\alpha\beta} + e^\alpha{}_{\dot\beta}
\eta^{\dot\alpha\dot\beta} + \frac{a_{3}}{6} e_{\beta\dot\beta}
\xi^{\alpha\beta\dot\alpha\dot\beta} - \frac{a_0}{4}
e^{\alpha\dot\alpha}\xi^{} \nonumber
\\
\delta B^{\alpha(2)} &=& \frac{a_0}{2} \eta^{\alpha(2)}, \qquad
\delta A = D\xi - \frac{a_0}{2} e_{\alpha\dot\alpha}
\xi^{\alpha\dot\alpha}\nonumber
\\
\delta \pi^{\alpha\dot\alpha} &=& -
\frac{a_0\tilde{a}_0}{24}\xi^{\alpha\dot\alpha},\qquad \delta\varphi
= \frac{\tilde{a}_0}{12} \nonumber
\end{eqnarray}
Compared to the massless case in the previous section, one can see
that we still have all the gauge symmetries that our massless fields
possessed modified so as to be consistent with the structure of the
massive Lagrangian. Such gauge invariant formulation of the massive
theory in $(A)dS_4$ space possesses some remarkable features.
Firstly, we can consider a flat limit $\lambda\rightarrow0$ and
immediately obtain the description of the massive fields in
Minkowski space. Secondly, in anti-de Sitter space when
$\lambda^2>0$ there is a correct massless limit $m\rightarrow0$
without the gap in the number of physical degrees of freedom. In
such a limit our system decomposes into two systems describing the
massless spin-$s$ and the massive spin-$(s-1)$ fields. Lastly, in de
Sitter space when $\lambda^2<0$ one can consider the so-called
partially massless limits $a_k\rightarrow0$. In such a limit, the
system decomposes into the two subsystems describing the partially
massless spin-$s$ field and the massive spin-$k$ field.

As in the massless case, to construct a complete set of the gauge
invariant objects one has to introduce a lot of extra fields which
do not, however, enter the free Lagrangian. In the following, we
restrict ourselves to the curvatures for the physical and auxiliary
fields only. With the explicit expressions for the gauge
transformations at our disposal (\ref{MasvBosonGT}), it is rather
straightforward to obtain (we omit all terms with the extra fields):
\begin{eqnarray}\label{MasvBosonCurv}
{\cal T}^{\alpha(k-1)\dot\alpha(k-1)} &=& D
f^{\alpha(k-1)\dot\alpha(k-1)} +
e_\beta{}^{\dot\alpha}\Omega^{\alpha(k-1)\beta\dot\alpha(k-2)} +
e^\alpha{}_{\dot\beta}\Omega^{\alpha(k-2)\dot\alpha(k-1)\dot\beta}
\nonumber
\\
 && + \frac{(k-1)a_{k+1}}{2(k+1)} e_{\beta\dot\beta}
f^{\alpha(k-1)\beta\dot\alpha(k-1)\dot\beta} + \frac{a_k}{2k(k-1)}
e^{\alpha\dot\alpha} f^{\alpha(k-2)\dot\alpha(k-2)} \nonumber
\\
{\cal R}^{\alpha(k),\dot\alpha(k-2)} &=& D
\Omega^{\alpha(k),\dot\alpha(k-2)} + \frac{a_{k+1}}{2}
e_{\beta\dot\beta} \Omega^{\alpha(k)\beta\dot\alpha(k-2)\dot\beta} +
\frac{a_k}{2k(k+1)} e^{\alpha\dot\alpha}
\Omega^{\alpha(k-1)\dot\alpha(k-3)} \nonumber
\\
 && + \frac{b_k}{2k} e^\alpha{}_{\dot\beta}
f^{\alpha(k-1)\dot\alpha(k-2)\dot\beta} \nonumber
\\
{\cal R}^{\alpha(2)} &=& D \Omega^{\alpha(2)} + \frac{a_3}{2}
e_{\beta\dot\beta} \Omega^{\alpha(2)\beta\dot\beta} + \frac{b_2}{4}
e^\alpha{}_{\dot\beta} f^{\alpha\dot\beta} - \frac{a_0}{4}
E^\alpha{}_\beta B^{\alpha\beta} +
\frac{a_0\tilde{a}_0}{24}E^{\alpha(2)} \varphi \nonumber
\\
{\cal T}^{\alpha\dot\alpha} &=& Df^{\alpha\dot\alpha} +
e_\beta{}^{\dot\alpha} \Omega^{\alpha\beta} + e^\alpha{}_{\dot\beta}
\Omega^{\dot\alpha\dot\beta} + \frac{a_{3}}{6} e_{\beta\dot\beta}
f^{\alpha\beta\dot\alpha\dot\beta} - \frac{a_0}{4}
e^{\alpha\dot\alpha} A
\\
{\cal C}^{\alpha(2)} &=& DB^{\alpha(2)} -
\frac{a_0}{2}\Omega^{\alpha(2)} - \frac{\tilde{a}_0}{24}
e^\alpha{}_{\dot\beta}\pi^{\alpha\dot\beta} \nonumber
\\
{\cal R} &=& DA + 2(E_{\alpha(2)} B^{\alpha(2)} +
E_{\dot\alpha(2)}B^{\dot\alpha(2)}) - \frac{a_0}{2}
e_{\alpha\dot\alpha} f^{\alpha\dot\alpha} \nonumber
\\
{\cal C}^{\alpha\dot\alpha} &=& D\pi^{\alpha\dot\alpha}
+\frac{a_0\tilde{a}_0}{24} f^{\alpha\dot\alpha} -
\frac{\tilde{a}_0}{12} (e_\beta{}^{\dot\alpha} B^{\alpha\beta} +
e^\alpha{}_{\dot\beta} B^{\dot\alpha\dot\beta}) +
\frac{a_0{}^2}{8}e^{\alpha\dot\alpha} \varphi \nonumber
\\
{\cal C} &=& D\varphi + e_{\alpha\dot\alpha} \pi^{\alpha\dot\alpha}
-\frac{\tilde{a}_0}{12} A \nonumber
\end{eqnarray}

In our construction of the massive supermultiplets we will consider
supertransformations for the physical fields only. However, in all
calculations we will heavily use the auxiliary fields equations
(on-shell conditions) as well as corresponding algebraic identities:
\begin{eqnarray}
{\cal T}^{\alpha(k-1)\dot\alpha(k-1)} \approx 0 &\Rightarrow&
e_\beta{}^{\dot\alpha} {\cal R}^{\alpha(k-1)\beta\dot\alpha(k-2)} +
e^\alpha{}_{\dot\beta} {\cal
R}^{\alpha(k-2)\dot\alpha(k-1)\dot\beta}\approx 0 \nonumber
\\
{\cal R} \approx 0 &\Rightarrow& E_{\alpha(2)}{\cal C}^{\alpha(2)}+
E_{\dot\alpha(2)} {\cal C}^{\dot\alpha(2)} \approx 0
\label{MasvOn-Shell}
\\
{\cal C} \approx 0 &\Rightarrow& e_{\alpha\dot\alpha} {\cal
C}^{\alpha\dot\alpha} \approx 0 \quad\Rightarrow\quad
E^{\alpha}{}_{\dot\gamma} {\cal C}^{\beta\dot\gamma} \approx \frac12
\varepsilon^{\alpha\beta} E_{\gamma}{}_{\dot\gamma} {\cal
C}^{\gamma\dot\gamma} \nonumber
\end{eqnarray}
The variation of the Lagrangian (\ref{MasvBosonLag}) under the
arbitrary variations for the physical fields takes the simple form
\begin{eqnarray}\label{MasvBosonVar}
\delta {\cal L} &=& - 2i\sum_{k=2}^s(-1)^k {\cal
R}^{\alpha(k-1)\beta\dot\alpha(k-2)} e_\beta{}^{\dot\beta} \delta
f_{\alpha(k-1)\dot\alpha(k-2)\dot\beta} \nonumber
\\
 && - 2i E_{\alpha(2)} {\cal C}^{\alpha(2)} \delta A + 12i
E_{\alpha\dot\alpha} {\cal C}^{\alpha\dot\alpha} \delta \varphi +
h.c.
\end{eqnarray}
Let us stress once again that this expression is such that all extra
fields drop out.

\subsection{Half-integer spin $s+1/2$}\label{MasbFerm}

In the gauge invariant formalism, the massive half-integer
spin-$s+1/2$ field is described by the set of massless fields with
spins $1/2\leq k+1/2\leq s+1/2$. Frame-like formulation for the
massless fermionic fields with spins ($k\geq1$) were considered
above, they are described by the one-forms
$(\Phi^{\alpha(k)\dot\alpha(k-1)},h.c.)$, while massless spin-$1/2$
is described by a physical zero-form $(\phi^\alpha,h.c.)$. The
Lagrangian for free massive field in $AdS_4$ have the form
\begin{eqnarray}\label{MasvFermLag}
{\cal L} &=& \sum_{k=1}^{s} (-1)^k
\Phi_{\alpha(k-1)\beta\dot\alpha(k-1)} e^\beta{}_{\dot\beta}
D\Phi^{\alpha(k-1)\dot\alpha(k-1)\dot\beta} - \phi_\alpha
E^\alpha{}_{\dot\alpha} D\phi^{\dot\alpha} \nonumber
\\
 && + \sum_{k=2}^{s} (-1)^k c_k [E^{\beta(2)}
\Phi_{\alpha(k-2)\beta(2)\dot\alpha(k-1)}
\Phi^{\alpha(k-2)\dot\alpha(k-1)}] + c_0 \Phi_\alpha
E^\alpha{}_{\dot\alpha} \phi^{\dot\alpha} + h.c. \nonumber
\\
 && + \sum_{k=1}^s (-1)^k d_{k}[(k+1)
\Psi_{\alpha(k-1)\beta\dot\alpha(k-1)} E^\beta{}_{\gamma}
\Psi^{\alpha(k-1)\gamma\dot\alpha(k-1)} \nonumber
\\
 && - (k-1)\Psi_{\alpha(k)\dot\alpha(k-2)\dot\beta}
E^{\dot\beta}{}_{\dot\gamma}
\Psi^{\alpha(k)\dot\alpha(k-2)\dot\gamma}] + 2d_1 E\phi_\alpha
\phi^\alpha + h.c.
\end{eqnarray}
where
\begin{eqnarray}
d_k &=& \pm \frac{(s+1)}{2k(k+1)}M_1, \qquad M_1{}^2 =
m_1{}^2+s^2\lambda^2 \nonumber
\\
c_k{}^2 &=& \frac{(s-k+1)(s+k+1)}{k^2}[M_1{}^2-k^2\lambda^2],
\label{fermion_data}
 \\
c_0{}^2 &=& 2s(s+2)[M_1{}^2-\lambda^2] \nonumber
\end{eqnarray}
In the following, we assume that the parameters $c_k,c_0,M_1$ are
positive.

Explicit form of the coefficients (\ref{fermion_data}) are
determined by the invariance of the Lagrangian under the following
gauge transformations
\begin{eqnarray}
\delta \Phi^{\alpha(k)\dot\alpha(k-1)} &=& D
\xi^{\alpha(k)\dot\alpha(k-1)} + e_\beta{}^{\dot\alpha}
\eta^{\alpha(k)\beta\dot\alpha(k-2)} + 2d_{k} e^\alpha{}_{\dot\beta}
\xi^{\alpha(k-1)\dot\alpha(k-1)\dot\beta} \nonumber
\\
 && + c_{k+1} e_{\beta\dot\beta}
\xi^{\alpha(k)\beta\dot\alpha(k-1)\dot\beta} +
\frac{c_k}{(k-1)(k+1)}e^{\alpha\dot\alpha}
\xi^{\alpha(k-1)\dot\alpha(k-2)}
\\
\delta \phi^\alpha &=& c_0\xi^\alpha \nonumber
\end{eqnarray}
The general structure of the Lagrangian (\ref{MasvFermLag}) is the
same as in the bosonic case. The first line is the sum of kinetic
terms, the second line contains cross-terms and the last two lines
are mass terms. In such a formulation we can take the correct
massless limit $m_1\rightarrow0$ in $AdS$ ($\lambda^2>0$) and the
correct partially massless limits $c_k\rightarrow0$ in $dS$
($\lambda^2<0$). Taking a flat limit $\lambda\rightarrow0$ we obtain
the description of the massive fermionic fields in Minkowski space.

As in the bosonic case, we restrict ourselves with the gauge
invariant curvatures for the physical fields only omitting all the
extra fields:
\begin{eqnarray}\label{MasvFermCurv}
{\cal F}^{\alpha(k)\dot\alpha(k-1)} &=& D
\Phi^{\alpha(k)\dot\alpha(k-1)} +2d_{k}
e^\alpha{}_{\dot\beta}\Phi^{\alpha(k-1)\dot\alpha(k-1)\dot\beta}
\nonumber
\\
 && + c_{k+1} e_{\beta\dot\beta}
\Phi^{\alpha(k)\beta\dot\alpha(k-1)\dot\beta} +
\frac{c_k}{(k-1)(k+1)}e^{\alpha\dot\alpha}
\Phi^{\alpha(k-1)\dot\alpha(k-2)} \nonumber
\\
{\cal F}^{\alpha} &=& D\Phi^{\alpha} + 2d_{1}
e^\alpha{}_{\dot\beta}\Phi^{\dot\beta} + c_2 e_{\beta\dot\beta}
\Phi^{\alpha\beta\dot\beta} - \frac{c_0}{3} E^\alpha{}_\beta
\phi^\beta
\\
{\cal C}^\alpha &=& D\phi^{\alpha} - c_0 \Phi^{\alpha} + 2d_1
e^{\alpha}{}_{\dot\beta} \phi^{\dot\beta} \nonumber
\end{eqnarray}
The variation of the Lagrangian (\ref{MasvFermLag}) under the
arbitrary variations of the physical fields has the following form
\begin{equation}\label{MasvFermVar}
\delta {\cal L} = - \sum_{k=1}^s(-1)^k {\cal
F}_{\alpha(k-1)\beta\dot\alpha(k-1)} e^\beta{}_{\dot\beta} \delta
\Psi^{\alpha(k-1)\dot\alpha(k-1)\dot\beta} - {\cal C}_\alpha
E^\alpha{}_{\dot\alpha} \delta \phi^{\dot\alpha} + h.c.
\end{equation}

\section{Massive higher spin superblocks}\label{Section3}

There are two types of massive $N=1$ supermultiplets, each one
containing two massive bosonic fields (with opposite parities) and
two massive fermionic ones:
$$
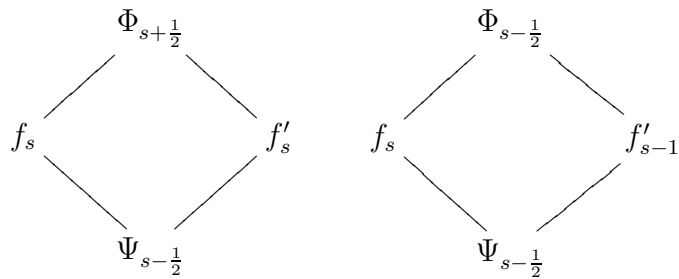
\xymatrix{  & \Phi_{s+\frac12} \ar@{-}[dr] &  \\
f_s \ar@{-}[ur] & & f'_s \ar@{-}[dl] \\
 & \Psi_{s-\frac12} \ar@{-}[ul] } \qquad
\xymatrix{  & \Phi_{s-\frac12} \ar@{-}[dr] &  \\
f_s \ar@{-}[ur] & & f'_{s-1} \ar@{-}[dl] \\
 & \Psi_{s-\frac12} \ar@{-}[ul] }
$$
To provide an explicit realization of such supermultiplets one has
to find supertransformations connecting each bosonic field with each
fermionic field so that: 1) the sum of the four free Lagrangians for
these fields is invariant; 2) the algebra of the
supertransformations is closed. In this work we use the following
strategy. Firstly, for each pair of bosonic and fermionic fields (we
call it superblock in what follows) we find the supertransformations
leaving the sum of their two Lagrangians invariant. Then we combine
all four fields together and adjust parameters of these superblocks
so that the algebra of the supertransformations is closed. One can
see from the diagrams above that there are only two non-trivial
superblocks, namely $(s,s+1/2)$ and $(s-1/2,s)$. Such a strategy
therefore greatly simplifies the whole construction.

In the gauge invariant formalism the description of massive higher
spin fields is constructed out of the appropriately chosen set of
massless ones. It seems natural to expect that one can construct a
description of massive higher spin supermultiplet out of an
appropriately chosen set of massless ones. Indeed, if one decomposes
all four massive fields into their massless components, the
resulting spectrum of massless components does correspond to some
set of massless supermultiplets. However, the explicit structure of
the supertransformations (see below) shows that all massless
components still remain connected with all their neighbours so that
the whole system looks just like one big massless supermultiplet
(similarly to what we obtained in the three dimensional case
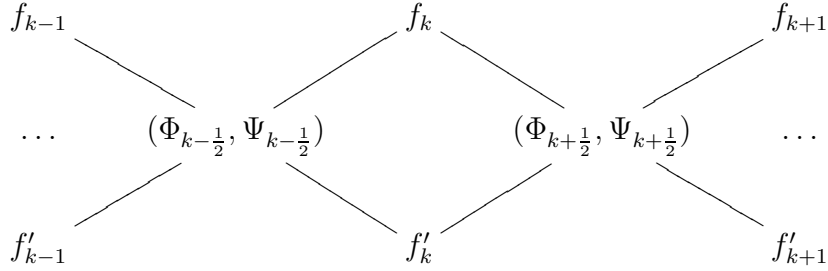
\cite{BSZ17}):
$$
\xymatrix{ f_{k-1} \ar@{-}[dr] & & f_k \ar@{-}[dr] & & f_{k+1} \\
\dots & (\Phi_{k-\frac12},\Psi_{k-\frac12}) \ar@{-}[ur] \ar@{-}[dr]
& &
(\Phi_{k+\frac12},\Psi_{k+\frac12}) \ar@{-}[ur] \ar@{-}[dr] & \dots \\
f'_{k-1} \ar@{-}[ur] & & f'_k \ar@{-}[ur] & & f'_{k+1} }
$$
One can introduce new fermionic variables:
\begin{eqnarray*}
\tilde{\Phi}_k &=& cos \Theta_k \Phi_k + \sin \Theta_k \Psi_k \\
\tilde{\Psi}_k &=& - sin \Theta_k \Phi_k + \cos \Theta_k \Psi_k
\end{eqnarray*}
and adjust mixing angles $\Theta_k$ so that the whole system
decomposes into the sum of massless supermultiplets containing two
bosonic and two fermionic fields:
$$
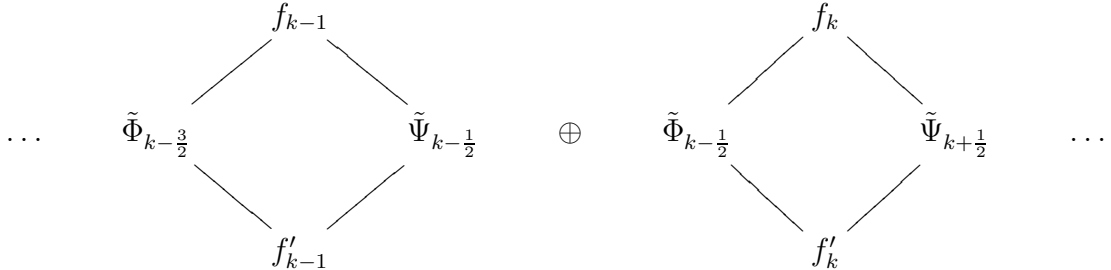
\xymatrix{ &  & f_{k-1} \ar@{-}[dr] & & & & f_k \ar@{-}[dr] & & \\
\dots & \tilde{\Phi}_{k-\frac32} \ar@{-}[ur] \ar@{-}[dr] & &
\tilde{\Psi}_{k-\frac12} & \oplus & \tilde{\Phi}_{k-\frac12}
\ar@{-}[ur] \ar@{-}[dr] & & \tilde{\Psi}_{k+\frac12} & \dots \\
 & & f'_{k-1} \ar@{-}[ur] & & & & f'_k \ar@{-}[ur] & & }
$$
The separation of these supermultiplets into the usual pairs is
impossible because the bosonic fields have opposite parities.
However in this case the structure of cross and mass-like terms in
the fermionic Lagrangian cease to be diagonal making the
construction of massive supermultiplets more complicated. Note that
it is this approach that was used in the previous works of one of
the current authors \cite{Zin07a}.

\subsection{Supertransformations}

We begin with a general discussion valid for the construction of
both massive superblocks and consider the most general ansatz for
the supertransformations. For the bosonic field variables we choose
\begin{eqnarray}\label{MasvSTevBos}
\delta f^{\alpha(k-1)\dot\alpha(k-1)} &=&
\alpha_{k-1}\Phi^{\alpha(k-1)\beta\dot\alpha(k-1)} \zeta_\beta -
\bar\alpha_{k-1}\Phi^{\alpha(k-1)\dot\alpha(k-1)\dot\beta}
\zeta_{\dot\beta} \nonumber\\
 && + \alpha'_{k-1} \Phi^{\alpha(k-1)\dot\alpha(k-2)}
\zeta^{\dot\alpha} - \bar\alpha'_{k-1}
\Phi^{\alpha(k-2)\dot\alpha(k-1)} \zeta^{\alpha} \nonumber
\\
\delta A &=& \alpha_0 \Phi^\alpha\zeta_\alpha - \bar\alpha_0
\Phi^{\dot\alpha} \zeta_{\dot\alpha} + \alpha'_0
e_{\alpha\dot\alpha}\psi^\alpha\zeta^{\dot\alpha} - \bar\alpha'_0
e_{\alpha\dot\alpha}\psi^{\dot\alpha} \zeta^{\alpha}
\\
\delta \varphi &=& \tilde\alpha_0 \phi^\alpha \zeta_{\alpha} -
\bar{\tilde\alpha}_0 \phi^{\dot\alpha} \zeta_{\dot\alpha} \nonumber
\end{eqnarray}
and for the fermionic ones
\begin{eqnarray}\label{MasvSTevFerm}
\delta \Phi^{\alpha(k)\dot\alpha(k-1)} &=&
\beta_{k-1}\Omega^{\alpha(k)\dot\alpha(k-2)} \zeta^{\dot\alpha} +
\gamma_{k-1}f^{\alpha(k-1)\dot\alpha(k-1)} \zeta^{\alpha} \nonumber
\\
 && + \beta'_{k-1} \Omega^{\alpha(k-1)\beta\dot\alpha(k-2)}
\zeta_\beta + \gamma'_{k-1}
f^{\alpha(k-1)\dot\alpha(k-2)\dot\beta}\zeta_{\dot\beta} \nonumber
\\
\delta \Phi^\alpha &=& \beta_0 e_{\beta\dot\beta}
B^{\alpha\beta}\zeta^{\dot\beta} + \beta'_{1} \Omega^{\alpha\beta}
\zeta_\beta +\gamma_0 A \zeta^\alpha +
\gamma'_{1}f^{\alpha\dot\beta} \zeta_{\dot\beta} + \hat\gamma_0
e^\alpha{}_{\dot\alpha} \varphi\zeta^{\dot\alpha}
\\
\delta \phi^\alpha &=& \tilde\beta_0 \pi^{\alpha\dot\alpha}
\zeta_{\dot\alpha} + \beta'_0 B^{\alpha\beta} \zeta_\beta +
\tilde\gamma_0 \varphi \zeta^\alpha \nonumber
\end{eqnarray}
where all coefficients are complex. One can see that the
supertransformations for higher spin components are combinations of
the massless supertransformations (\ref{MaslST1}) and
(\ref{MaslST2}).
$$
\xymatrix{ f_{k-1} \ar@{-}[dr]^-{\alpha_{k-1}} & & f_k
\ar@{-}[dr]^-{\alpha_k} & & f_{k+1} \\
\dots & \Phi_{k-\frac12} \ar@{-}[ur]^-{\alpha'_k} & &
\Phi_{k+\frac12} \ar@{-}[ur]^{\alpha'_{k+1}} & \dots }
$$
The ansatz for the supertransformations (\ref{MasvSTevBos}),
(\ref{MasvSTevFerm}) has the same form for both massive superblocks
$(s+1/2,s)$ and $(s,s-1/2)$, the only difference being in the
boundary conditions. In the first case we have
\begin{equation}\label{InitCond1}
\alpha_s = \beta_s = \gamma_s = 0, \qquad \alpha'_s = \beta'_s =
\gamma'_s = 0
\end{equation}
while in the second case
\begin{equation}\label{InitCond2}
\alpha_{s-1} = \beta_{s-1} = \gamma_{s-1} = 0, \qquad \alpha'_s =
\beta'_s = \gamma'_s = 0
\end{equation}

The variation of the sum of the bosonic and fermionic Lagrangians
(\ref{MasvBosonVar}), (\ref{MasvFermVar}) under the
supertransformations (\ref{MasvSTevBos}), (\ref{MasvSTevFerm}) has
the form $\delta{\cal L}+\delta{\cal L}'$, where
\begin{eqnarray}\label{MasvVar}
\delta {\cal L} &=& (-1)^k \sum_{k=2}^{s}[- (k-1)\bar\beta_{k-1}
{\cal F}_{\alpha(k-1)\beta\dot\alpha(k-1)} e^\beta{}_{\dot\beta}
\Omega^{\alpha(k-2)\dot\alpha(k-1)\dot\beta} \zeta^{\alpha}
\nonumber
\\
 && + 4i\alpha_{k-1} \Phi_{\alpha(k-2)\beta\gamma\dot\alpha(k-1)}
e^\gamma{}_{\dot\gamma} {\cal
R}^{\alpha(k-2)\dot\alpha(k-1)\dot\gamma} \zeta^{\beta} \nonumber
\\
 && - \bar\gamma_{k-1} {\cal F}_{\alpha(k-1)\beta\dot\alpha(k-1)}
e^\beta{}_{\dot\beta} (f^{\alpha(k-1)\dot\alpha(k-1)}
\zeta^{\dot\beta} +
(k-1)f^{\alpha(k-1)\dot\alpha(k-2)\dot\beta}\zeta^{\dot\alpha})]
\nonumber
\\
 && - \bar\beta_0 E_{\dot\alpha\dot\beta} {\cal F}_{\alpha}
B^{\dot\alpha\dot\beta} \zeta^{\alpha} + 4i\alpha_0
E_{\dot\beta(2)}\Phi_{\alpha} {\cal C}^{\dot\beta(2)} \zeta^{\alpha}
\nonumber
\\
 && + \bar{\tilde\beta}_0 E^\alpha{}_{\dot\alpha} {\cal C}_\alpha
\pi^{\beta\dot\alpha} \zeta_{\beta} + 12i\tilde\alpha_0
E_{\beta\dot\beta} \phi^\alpha {\cal C}^{\beta\dot\beta}
\zeta_{\alpha} \nonumber
\\
 && + \bar\gamma_0 e^\alpha{}_{\dot\alpha} {\cal F}_{\alpha}
A\zeta^{\dot\alpha} + \bar{\tilde\gamma}_0 E^\alpha{}_{\dot\alpha}
{\cal C}_\alpha \varphi \zeta^{\dot\alpha} + 2\bar{\hat\gamma}_0
E^\alpha{}_\beta {\cal F}_{\alpha} \varphi \zeta^{\beta} + h.c.
\nonumber
\\
\delta {\cal L}' &=& \sum_{k=2}^{s} (-1)^k [\bar\beta'_{k-1} {\cal
F}_{\alpha(k-2)\gamma\dot\alpha(k-2)} e^\gamma{}_{\dot\gamma}
\Omega^{\alpha(k-2)\dot\alpha(k-2)\dot\gamma\dot\beta}
\zeta_{\dot\beta} \nonumber
\\
 && - 4i(k-1)\alpha'_{k-1} \Phi_{\alpha(k-2)\beta\dot\alpha(k-2)}
e^\beta{}_{\dot\beta} {\cal
R}^{\alpha(k-2)\dot\alpha(k-2)\dot\beta\dot\gamma}
\zeta_{\dot\gamma} \nonumber
\\
 && + \bar\gamma'_{k-1} {\cal F}_{\alpha(k-2)\gamma\dot\alpha(k-2)}
e^\gamma{}_{\dot\gamma}
f^{\alpha(k-2)\beta\dot\alpha(k-2)\dot\gamma}\zeta_\beta] \nonumber
\\
 && - \bar\beta'_0 {\cal C}_\alpha E^\alpha{}_{\dot\alpha}
B^{\dot\alpha\dot\beta} \zeta_{\dot\beta} + 8i \alpha'_0
\psi_{\alpha}E^\alpha{}_{\dot\beta} {\cal C}^{\dot\beta(2)}
\zeta_{\dot\beta} + h.c.
\end{eqnarray}
Here we used the equations for the auxiliary bosonic fields
(\ref{MasvOn-Shell}). We now proceed as in the massless case,
deriving the corresponding identities. Let us recall a general
scheme. Lagrangian variation (\ref{MasvVar}) has the structure
$$
\delta{\cal L} = ({\cal F}\Omega|_\beta \oplus \Phi {\cal R}|_\alpha
\oplus {\cal F}f|_\gamma) \zeta
$$
where $\Phi$, ${\cal F}$ are sets of all fields and curvatures for
the fermion, $f$ is a set of physical fields for the boson and
$\Omega$, ${\cal R}$ are sets of auxiliary fields and curvatures for
the boson. Since a Lagrangian is defined up to a total derivative we
have two type of identities $ D[\Phi\Omega\zeta]=0$, $D[\Phi
f\zeta]=0$. They lead to
\begin{eqnarray}
\label{SchemIden1} {\cal F} \Omega \zeta &=& \Phi {\cal R} \zeta
\oplus \Delta (\Phi\Omega,\Phi f) \zeta
\\
\label{SchemIden2} {\cal F} f \zeta &=& \Phi {\cal T} \zeta \oplus
\Delta (\Phi\Omega,\Phi f) \zeta
\end{eqnarray}
Using the explicit form of identities (\ref{SchemIden1}),
(\ref{SchemIden2}) (see Appendix A for details), we obtain
expressions for the parameters $\alpha$ and $\gamma$ in terms of
$\beta$:
\begin{eqnarray}
\alpha_{k-1} &=& i\frac{(k-1)}{4} \bar\beta_{k-1}, \qquad
\alpha'_{k-1} = \frac{i}{4(k-1)} \bar\beta'_{k-1} \nonumber
\\
\alpha_0 &=& -i\frac{\bar\beta_0}{4}, \qquad \tilde\alpha_0 =
i\frac{\bar{\tilde\beta}_0}{24}, \qquad \alpha'_0 = -
i\frac{\bar\beta'_0}{8} \label{SolutAlf}
\end{eqnarray}
and
\begin{eqnarray}
\gamma_{k-1} &=& 2d_k \bar\beta_{k-1} ,\qquad \gamma'_{k-1} =
2d_{k-1} \bar\beta'_{k-1} \nonumber
\\
\gamma_ 0 &=& - d_1 \bar\beta_0, \quad \tilde\gamma_0 = -
12\frac{c_0d_1}{\tilde{a}_0} \bar\beta_0, \quad \hat\gamma_0 = -
\frac18 \tilde{a}_0 \beta_0 \label{SolutGam}
\end{eqnarray}
and we also obtain recurrent equations on the parameters $\beta_k$
\begin{eqnarray}
2(k+1)\beta_{k-1} c_{k+1} &=& k\beta_{k} a_{k+1}, \qquad 2c_2
\beta_0 = a_0\beta_1, \qquad 2c_0\tilde\beta_0 = - {\tilde{a}_0}
\beta_0 \label{RecurEq1}
\\
\frac12 \beta'_{k-1} a_{k+1} &=& \beta'_kc_k, \qquad \frac12 a_0
\beta'_0 = c_0 \beta'_1, \qquad {\tilde{a}_0} \bar{\tilde\beta}_0 =
- 24d_1 \beta'_0 \label{RecurEq2}
\end{eqnarray}
as well as four independent equations which relate $\beta$ and
$\beta'$ and the bosonic and fermionic mass parameters:
\begin{eqnarray}
0 &=& \frac{\beta'_{k-1}c_k}{(k-1)} - \frac{\beta'_ka_{k+1}}{2(k+1)}
+ \lambda \beta_{k-1} - \gamma_{k-1} \label{Relation1}
\\
0 &=& (k-1)\beta_{k-1}c_{k} - \frac{(k-2)}{2} \beta_{k-2}a_{k} -
\lambda \beta'_{k-1} + \gamma'_{k-1} \label{Relation2}
\\
0 &=& \frac{(k-1)}{2k} \bar\beta_{k-1}b_{k} - 2kd_{k}\gamma_{k-1}
+\lambda \bar\gamma_{k-1} -
\frac{\bar\gamma'_{k}a_{k+1}}{2k(k+1)}\label{Relation3}
\\
0 &=& \frac{\bar\beta'_{k-1}b_{k}}{2} - 2(k-1)d_{k-1}\gamma'_{k-1}
-\lambda\bar\gamma'_{k-1} + \bar\gamma_{k-1}c_k \label{Relation4}
\end{eqnarray}
The explicit solution of these equations depends on the concrete
massive superblock. More spefically, it depends on the initial
conditions (\ref{InitCond1}), (\ref{InitCond2}) and on the sign of
$d_k$, i.e. on the sign before massive terms in Lagrangian for
fermions. In the following we present exact solutions for two
massive superblocks $(s+1/2,s)$ and $(s,s-1/2,)$.

\subsection{Superblock $(s+1/2,s)$}\label{MasvSB1}

Here we present our results for the massive superblock $(s+1/2,s)$.
The massive boson spin-$s$ with the mass parameter $M$, as described
in section \ref{MasbBoson}, we have
\begin{eqnarray*}
b_k &=& \frac{2s(s+1)}{k(k-1)(k+1)}M^2, \qquad m{}^2 = M{}^2 -
s(s-1)\lambda^2 \nonumber
\\
a_k{}^2 &=& \frac{4(s-k+1)(s+k)}{(k-1)(k-2)} [M^2 -
k(k-1)\lambda^2]\nonumber
\\
a_0{}^2 &=& 2(s-1)(s+2)[M^2 - 2\lambda^2], \qquad \tilde{a}_0{}^2 =
24s(s+1)M^2
\end{eqnarray*}
The massive fermion spin-$(s+1/2)$ with mass parameter $M_1$ as
described in section \ref{MasbFerm}, here
\begin{eqnarray*}
d_k &=& \pm\frac{(s+1)}{2k(k+1)}M_1, \qquad m_1{}^2 =
M_1{}^2-s^2\lambda^2 \nonumber
\\
c_k{}^2 &=& \frac{(s-k+1)(s+k+1)}{k^2} [M_1{}^2 - k^2\lambda^2]
\\
c_0{}^2 &=& 2s(s+2) [M_1{}^2 - \lambda^2]
\end{eqnarray*}
Supertransformations for massive superblock $(s+1/2,s)$ have the
form (\ref{MasvSTevBos}), (\ref{MasvSTevFerm}) with the initial
conditions (\ref{InitCond1}). The parameters $\alpha_k$ and
$\gamma_k$ are determined by (\ref{SolutAlf}) and (\ref{SolutGam}).
From the equation (\ref{Relation3}) one can obtain an important
relation on the bosonic and fermionic mass parameters. Indeed, at
$k=s$ we have
$$
(M^2-M_1{}^2) \bar\beta_{s-1} = \mp M_1\lambda \beta_{s-1}
$$
where the sign corresponds to that of $d_k$. So we have four
independent cases
\begin{eqnarray*}
M^2 &=& M_1(M_1-\lambda), \qquad \bar\beta_{s-1} = \pm\beta_{s-1}
\\
M^2 &=& M_1(M_1+\lambda), \qquad \bar\beta_{s-1} = \mp \beta_{s-1}
\end{eqnarray*}
The solution of other equations give, for $M^2=M_1(M_1-\lambda)$
\begin{eqnarray*}
\beta_{k-1} &=& \sqrt{\frac{(s+k+1)(M_1+k\lambda)}{(k-1)}}\beta,
\quad \beta'_{k-1} = \sqrt{{(k-1)(s-k+1)(M_1-k\lambda)}}\beta
\\
\beta_{0} &=& \sqrt{{2(s+2)(M_1+\lambda)}}\beta, \quad \beta'_{0} =
2\sqrt{{s(M_1-\lambda)}}\beta, \quad \tilde\beta_0 = -
\sqrt{6(s+1)M_1}\beta
\end{eqnarray*}
and for $M^2=M_1(M_1+\lambda)$
\begin{eqnarray*}
\beta_{k-1}{} &=&
\sqrt{\frac{(s+k+1)(M_1-k\lambda)}{(k-1)}}\beta,\quad \beta'_{k-1} =
- \sqrt{{(k-1)(s-k+1)(M_1+k\lambda)}}\beta
\\
\beta_{0} &=& \sqrt{{2(s+2)(M_1-\lambda)}}\beta, \quad \beta'_{0} =
- 2\sqrt{{s(M_1+\lambda)}}\beta, \quad \tilde\beta_0 = -
\sqrt{6(s+1)M_1}\beta
\end{eqnarray*}

Therefore, we see that in the two cases the parameters $\beta$ are
real and in the other two they are imaginary. This means that one
half of the solutions corresponds to the massive superblocks with
the parity-even boson while another half corresponds to the massive
superblocks with the parity-odd one.

In order to present these four cases for the massive superblock
$(s,s+1/2)$ in a more clear form let us introduce following
notations. We denote integer spin $s$ with the mass parameter $M$ as
\begin{equation}\label{NotatBos}
[s]_M^\pm
\end{equation}
here  $\pm$ corresponds to parity-even/parity-odd boson. We also
denote half-integer spin $s+1/2$ with the mass parameter $M_1$ as
\begin{equation}\label{NotatFer}
[s+\frac12]_{ M_1}^{\pm}
\end{equation}
here  $\pm$ corresponds to the sign of $d_k$. In these notations the
four solutions for the massive superblock given above look as
follows:
$$
1) \left( \begin{array}{c} {[s]}_{M}^+ \\
{[s+\frac12]}_{M_1}^+ \end{array} \right), \quad 2) \left(
\begin{array}{c} {[s]}_{M}^- \\ {[s+\frac12]}_{M_1}^{-}
\end{array} \right), \quad
3) \left( \begin{array}{c} {[s]}_{M'}^+ \\
{[s+\frac12]}_{M_1}^{-} \end{array} \right),\quad 4)\left(
\begin{array}{c} {[s]}_{M'}^- \\ {[s+\frac12]}_{M_1}^{+}
\end{array} \right)
$$
where
$$
M^2 = M_1(M_1-\lambda), \qquad M'^2 = M_1(M_1+\lambda)
$$
In the first and third cases we have
\begin{eqnarray*}
\beta_{k-1} &=& \sqrt{\frac{(s+k+1)(M_1\pm
k\lambda)}{(k-1)}}\rho,\quad \beta'_{k-1} =
\pm\sqrt{{(k-1)(s-k+1)(M_1\mp k\lambda)}}\rho
\\
\beta_{0} &=& \sqrt{{2(s+2)(M_1\pm\lambda)}}\rho, \quad \beta'_{0} =
\pm2\sqrt{{s(M_1\mp\lambda)}}\rho, \quad \tilde\beta_0 = -
\sqrt{6(s+1)M_1}\rho
\end{eqnarray*}
here the upper sign corresponds to 1) and the lower sign corresponds
to 3). In the second and fourth cases we have
\begin{eqnarray*}
\beta_{k-1} &=& i\sqrt{\frac{(s+k+1)(M_1\pm
k\lambda)}{(k-1)}}\rho,\quad \beta'_{k-1} = \pm
i\sqrt{{(k-1)(s-k+1)(M_1\mp k\lambda)}}\rho
\\
\beta_{0} &=& i\sqrt{{2(s+2)(M_1\pm\lambda)}}\rho, \quad \beta'_{0}
= \pm i2\sqrt{{s(M_1\mp\lambda)}}\rho, \quad \tilde\beta_0 = -
i\sqrt{6(s+1)M_1}\rho
\end{eqnarray*}
here the upper sign corresponds to 2) and the lower sign corresponds
to 4).

\subsection{Superblock $(s,s-1/2)$}\label{MasvSB2}

Here we collect our results for the massive superblock $(s,s-1/2)$.
For the massive even or odd spin-$s$ boson with the mass parameter
$M$ we use the same formulation as in the previous subsection, while
for the massive spin-$(s-1/2)$ fermion with the mass parameter $M_2$
we use its description in section \ref{MasbFerm} with the shift
$s\rightarrow(s-1)$:
\begin{eqnarray*}
d_k &=& \pm\frac{s}{2k(k+1)}M_2, \qquad m_2{}^2 =
M_2{}^2-(s-1)^2\lambda^2
\\
c_k{}^2 &=& \frac{(s-k)(s+k)}{k^2}[M_2{}^2-k^2\lambda^2]
\\
c_0{}^2 &=& 2(s-1)(s+1)[M_2{}^2-\lambda^2]
\end{eqnarray*}
Supertransformations for the massive superblock $(s-1/2,s)$ are the
same as in the previous case (\ref{MasvSTevBos}),
(\ref{MasvSTevFerm}) but with different initial conditions
(\ref{InitCond2}). The parameters $\alpha_k$ and $\gamma_k$ are
still determined by (\ref{SolutAlf}) and (\ref{SolutGam}). From the
equation (\ref{Relation2}) one can relate bosonic and fermionic mass
parameters, indeed at $k=s$ we have
$$
(M^2-M_2{}^2)\bar\beta'_{s-1} = \pm M_2\lambda\beta'_{s-1}
$$
here the sign corresponds to that of $d_k$. So we again have four
independent cases
\begin{eqnarray*}
M^2 &=& M_2(M_2+\lambda), \qquad \bar\beta'_{s-1} = \pm\beta'_{s-1}
\\
M^2 &=& M_2(M_2-\lambda), \qquad \bar\beta'_{s-1} = \mp\beta'_{s-1}
\end{eqnarray*}
The solution of other equations gives, for $M^2=M_2(M_2+\lambda)$
\begin{eqnarray*}
\beta_{k-1} &=& \sqrt{\frac{(s-k)(M_2-k\lambda)}{(k-1)}}\beta,
\qquad \beta'_{k-1} = \sqrt{(k-1)(s+k)(M_2+k\lambda)}\beta
\\
\beta_{0} &=& \sqrt{{2(s-1)(M_2-\lambda)}}\beta, \quad \beta'_{0} =
2\sqrt{{(s+1)(M_2+\lambda)}}\beta, \quad \tilde\beta_0 = -
\sqrt{6sM_2}\beta
\end{eqnarray*}
and for $M^2=M_2(M_2-\lambda)$
\begin{eqnarray*}
\beta_{k-1} &=& \sqrt{\frac{(s-k)(M_2+k\lambda)}{(k-1)}}\beta, \quad
\beta'_{k-1} = - \sqrt{(k-1)(s+k)(M_2-k\lambda)}\beta
\\
\beta_{0} &=& \sqrt{{2(s-1)(M_2+\lambda)}}\beta, \quad \beta'_{0} =
- 2\sqrt{{(s+1)(M_2-\lambda)}}\beta, \quad \tilde\beta_0 = -
\sqrt{6sM_2}\beta
\end{eqnarray*}
Again we see that in two cases the parameters $\beta$ are real and
in other two cases they are imaginary. They correspond to the
massive superblocks with the parity-even and parity-odd bosons
respectively. Using notations (\ref{NotatBos}), (\ref{NotatFer})
these four cases for the massive $(s,s-1/2)$ superblock can be
presented as
$$
1) \left( \begin{array}{c} {[s]}_{M}^+ \\
{[s-\frac12]}_{M_2}^+ \end{array} \right), \quad
2) \left( \begin{array}{c} {[s]}_{M}^- \\
{[s-\frac12]}_{M_2}^{-} \end{array} \right), \quad
3) \left( \begin{array}{c} {[s]}_{M'}^+ \\
{[s-\frac12]}_{M_2}^{-} \end{array} \right), \quad
4)\left( \begin{array}{c} {[s]}_{M'}^- \\
{[s-\frac12]}_{M_2}^{+} \end{array} \right)
$$
where
$$
M^2=M_2(M_2+\lambda),\qquad M'^2=M_2(M_2-\lambda)
$$
For the first and third cases we have
\begin{eqnarray*}
\beta_{k-1} &=& \sqrt{\frac{(s-k)(M_2\mp k\lambda)}{(k-1)}}\rho,
\quad \beta'_{k-1} = \pm\sqrt{(k-1)(s+k)(M_2\pm k\lambda)}\rho
\\
\beta_{0} &=& \sqrt{{2(s-1)(M_2\mp\lambda)}}\rho, \quad \beta'_{0} =
\pm2\sqrt{{(s+1)(M_2\pm\lambda)}}\rho, \quad \tilde\beta_0 = -
\sqrt{6sM_2}\rho
\end{eqnarray*}
here the upper sign corresponds to 1) and the lower sign corresponds
to 3). In the second and fourth cases we have
\begin{eqnarray*}
\beta_{k-1} &=& i\sqrt{\frac{(s-k)(M_2\mp k\lambda)}{(k-1)}}\rho,
\quad \beta'_{k-1} = \pm i\sqrt{(k-1)(s+k)(M_2\pm k\lambda)}\rho
\\
\beta_{0} &=& i\sqrt{{2(s-1)(M_2\mp\lambda)}}\rho, \quad \beta'_{0}
= \pm i2\sqrt{{(s+1)(M_2\pm\lambda)}}\rho, \quad \tilde\beta_0 = -
i\sqrt{6sM_2}\rho
\end{eqnarray*}
here the upper sign correspond to 2) and the lower sign correspond
to 4). In all four cases $\rho$ is real.

\section{Massive higher spin supermultiplets}\label{Section4}

In the previous section we constructed massive superblocks
containing one massive fermion and one massive boson. For each
individual superblock, we found supertransformations defined up to a
one common parameter $\rho$. In this section we use these results to
construct complete massive supermultiplets. For that we choose
appropriate solutions for each superblock and adjust their
parameters so that the algebra of these supertransformations is
closed. In the next subsection, we consider general properties of
such construction and then present our results for the case of
integer and half-integer superspins.

\subsection{General construction}

Any massive $N=1$ supermultiplet contains two massive fermions and
two massive bosons. In the notations given in the previous section
(\ref{NotatBos}), (\ref{NotatFer}) they have the following structure
$$
\xymatrix {& {[s+\frac12]_{M_1}^{+}} \ar@{-}[dl]_-{\rho_1}
\ar@{-}[dr]^-{\rho_3} &\\
 {[s]_{M}^+} \ar@{-}[dr]_-{\rho_2} & Y = s &
{[{s}]_{M'}^-} \\
 & {[s-\frac12]_{M_2}^-} \ar@{-}[ur]_-{\rho_4} & }
\qquad \xymatrix {& {[s-\frac12]_{M_1}^+} \ar@{-}[dl]_-{\rho_1}
\ar@{-}[dr]^-{\rho_3} &\\
 {[s]_{M}^+} \ar@{-}[dr]_-{\rho_2} & Y = s-\frac12 &
{[{s-1}]_{M'}^-}\\
 & {[s-\frac12]_{M_2}^-} \ar@{-}[ur]_-{\rho_4} & }
$$
As already mentioned, the two bosonic fields must have opposite
parities and it appears that the two fermionic fields must have
opposite signs of the mass terms. Let us introduce notations
$(f_+,\Omega_+)$ for the parity-even boson and $(f_-,\Omega_-)$ for
the parity-odd one. The fermions we denote as $\Phi_+,\Phi_-$
according to the sign of $d_k$.

The ansatz for the supertransformations is a combination of four
possible superblocks corresponding to the lines with the parameters
$\rho_{1,2,3,4}$. For example, for the parity-even boson we take:
\begin{eqnarray*}
\delta f_+^{\alpha(k-1)\dot\alpha(k-1)} &=&
\alpha_{k-1}|_{\rho_1}\Phi_+^{\alpha(k-1)\beta\dot\alpha(k-1)}
\zeta_\beta - \bar\alpha_{k-1}|_{\rho_1}
\Phi_+^{\alpha(k-1)\dot\alpha(k-1)\dot\beta} \zeta_{\dot\beta}
\\
 && + \alpha'_{k-1}|_{\rho_1}
\Phi_+^{\alpha(k-1)\dot\alpha(k-2)}\zeta^{\dot\alpha} -
\bar\alpha'_{k-1}|_{\rho_1} \Phi_+^{\alpha(k-2)\dot\alpha(k-1)}
\zeta^{\alpha}
\\
 && + \alpha_{k-1}|_{\rho_2}
\Phi_-^{\alpha(k-1)\beta\dot\alpha(k-1)}\zeta_\beta -
\bar\alpha_{k-1}|_{\rho_2}
\Phi_-^{\alpha(k-1)\dot\alpha(k-1)\dot\beta} \zeta_{\dot\beta}
\\
 && + \alpha'_{k-1}|_{\rho_2}
\Phi_-^{\alpha(k-1)\dot\alpha(k-2)}\zeta^{\dot\alpha} -
\bar\alpha'_{k-1}|_{\rho_2} \Phi_-^{\alpha(k-2)\dot\alpha(k-1)}
\zeta^{\alpha}
\\
\delta \Phi_+^{\alpha(k)\dot\alpha(k-1)} &=& \beta_{k-1}|_{\rho_1}
\Omega_+^{\alpha(k)\dot\alpha(k-2)} \zeta^{\dot\alpha} +
\gamma_{k-1}|_{\rho_1} f_+^{\alpha(k-1)\dot\alpha(k-1)}
\zeta^{\alpha}
\\
 && + \beta'_{k}|_{\rho_1}
\Omega_+^{\alpha(k)\beta\dot\alpha(k-1)}\zeta_\beta +
\gamma'_{k}|_{\rho_1} f_+^{\alpha(k)\dot\alpha(k-1)\dot\beta}
\zeta_{\dot\beta}
\\
\delta \Phi_-^{\alpha(k)\dot\alpha(k-1)} &=& \beta_{k-1}|_{\rho_2}
\Omega_+^{\alpha(k)\dot\alpha(k-2)} \zeta^{\dot\alpha} +
\gamma_{k-1}|_{\rho_2} f_+^{\alpha(k-1)\dot\alpha(k-1)}
\zeta^{\alpha}
\\
 && + \beta'_{k}|_{\rho_2}
\Omega_+^{\alpha(k)\beta\dot\alpha(k-1)} \zeta_\beta +
\gamma'_{k}|_{\rho_2} f_+^{\alpha(k)\dot\alpha(k-1)\dot\beta}
\zeta_{\dot\beta}
\end{eqnarray*}
(and similarly for the lower spin components), while the ansatz for
the parity-odd one can be obtained by replacement $\rho_1 \to
\rho_3$ and $\rho_2 \to \rho_4$.

The commutator of the two supertransformations must produce a
combination of translations and Lorentz transformations:
\begin{equation}\label{SA}
\{Q_\alpha,Q_{\dot\alpha}\}\sim P_{\alpha\dot\alpha},\quad
\{Q_\alpha,Q_{\alpha}\}\sim \lambda M_{\alpha\alpha},\quad
\{Q_{\dot\alpha},Q_{\dot\alpha}\}\sim \lambda
M_{\dot\alpha\dot\alpha}
\end{equation}
The structure of the mass-shell condition (\ref{MasvOn-Shell}) shows
that, for example, the commutator on the bosonic field
$f_+{}^{\alpha(k-1)\dot\alpha(k-1)}$ must contain fields
$\Omega_+{}^{\alpha(k)\dot\alpha(k-2)}$,
$\Omega_+{}^{\alpha(k-2)\dot\alpha(k)}$,
$f_+^{\alpha(k)\dot\alpha(k)}$, $f_+{}^{\alpha(k-1)\dot\alpha(k-1)}$
and $f_+{}^{\alpha(k-2)\dot\alpha(k-2)}$ only. This gives a number
of relations on the parameters:
$$
\alpha_{k-1}|_{\rho_1} \beta'_{k}|_{\rho_1} + \alpha_{k-1}|_{\rho_2}
\beta'_{k}|_{\rho_2} = 0, \qquad \alpha'_{k-1}|_{\rho_1}
\beta_{k-2}|_{\rho_1} + \alpha'_{k-1}|_{\rho_2}
\beta_{k-2}|_{\rho_2} = 0
$$
$$
\alpha_{k-1}|_{\rho_1} \beta'_{k}|_{\rho_3} + \alpha_{k-1}|_{\rho_2}
\beta'_{k}|_{\rho_4} = 0, \qquad \alpha'_{k-1}|_{\rho_1}
\beta_{k-2}|_{\rho_3} + \alpha'_{k-1}|_{\rho_2}
\beta_{k-2}|_{\rho_4} = 0
$$
$$
\alpha_{k-1}|_{\rho_1} \beta_{k-1}|_{\rho_3} +
\alpha'_{k-1}|_{\rho_1} \beta'_{k-1}|_{\rho_3} +
\alpha_{k-1}|_{\rho_2} \beta_{k-1}|_{\rho_4} +
\alpha'_{k-1}|_{\rho_2} \beta'_{k-1}|_{\rho_4} = 0
$$
$$
\alpha_{k-1}|_{\rho_1} \gamma_{k-1}|_{\rho_3} -
\bar\alpha'_{k-1}|_{\rho_1} \bar\gamma'_{k-1}|_{\rho_3} +
\alpha_{k-1}|_{\rho_2} \gamma_{k-1}|_{\rho_4} -
\bar\alpha'_{k-1}|_{\rho_2} \bar\gamma'_{k-1}|_{\rho_4} = 0
$$
$$
\alpha_{k-1}|_{\rho_1} \gamma'_{k}|_{\rho_3} -
\bar\alpha_{k-1}|_{\rho_1} \bar\gamma'_{k}|_{\rho_3} +
\alpha_{k-1}|_{\rho_2} \gamma'_{k}|_{\rho_4} -
\bar\alpha_{k-1}|_{\rho_2} \bar\gamma'_{k}|_{\rho_4} = 0
$$
$$
\alpha'_{k-1}|_{\rho_1} \gamma_{k-2}|_{\rho_3} -
\bar\alpha'_{k-1}|_{\rho_1} \bar\gamma_{k-2}|_{\rho_3} +
\alpha'_{k-1}|_{\rho_2} \gamma_{k-2}|_{\rho_4} -
\bar\alpha'_{k-1}|_{\rho_2} \bar\gamma_{k-2}|_{\rho_4} = 0
$$
If these relations are fulfilled the resulting expression for the
commutator has the form:
\begin{eqnarray*}
\ [\delta_1, \delta_2 ] f_+^{\alpha(k-1)\dot\alpha(k-1)} &=&
 (\alpha_{k-1}|_{\rho_1} \beta_{k-1}|_{\rho_1}
+ \alpha'_{k-1}|_{\rho_1} \beta'_{k-1}|_{\rho_1} +
\alpha_{k-1}|_{\rho_2} \beta_{k-1}|_{\rho_2} +
\alpha'_{k-1}|_{\rho_2} \beta'_{k-1}|_{\rho_2})
\\
 && \cdot [\Omega_+^{\alpha(k-1)\gamma\dot\alpha(k-2)}
(\zeta_1^{\dot\alpha} \zeta_{2\gamma} -
\zeta_2^{\dot\alpha}\zeta_{1\gamma}) +
\Omega_+^{\alpha(k-2)\dot\alpha(k-1)\dot\gamma} (\zeta_1^{\alpha}
\zeta_{2\dot\gamma} - \zeta_2^{\alpha} \zeta_{1\dot\gamma})]
\\
 && + (\alpha_{k-1}|_{\rho_1} \gamma'_k|_{\rho_1}
+ \alpha_{k-1}|_{\rho_2} \gamma'_k|_{\rho_2})
f_+^{\alpha(k-1)\gamma\dot\alpha(k-1)\dot\beta} (\zeta_{1\dot\beta}
\zeta_{2\gamma} - \zeta_{2\dot\beta}\zeta_{1\gamma})
\\
 && + (\alpha'_{k-1}|_{\rho_1} \gamma_{k-2}|_{\rho_1}
+ \alpha'_{k-1}|_{\rho_2} \gamma_{k-2}|_{\rho_2})
f_+^{\alpha(k-2)\dot\alpha(k-2)} (\zeta_1^{\alpha}
\zeta_2^{\dot\alpha} - \zeta_2^{\alpha} \zeta_1^{\dot\alpha})
\\
 && + (\alpha_{k-1}|_{\rho_1} \gamma_{k-1}|_{\rho_1}
 + \alpha'_{k-1}|_{\rho_1} \gamma'_{k-1}|_{\rho_1}
 + \alpha_{k-1}|_{\rho_2} \gamma_{k-1}|_{\rho_2}
 + \alpha'_{k-1}|_{\rho_2} \gamma'_{k-1}|_{\rho_2})
\\
 && \cdot [ f_+^{\alpha(k-2)\gamma\dot\alpha(k-1)}
(\zeta_1^{\alpha}\zeta_{2\gamma} - \zeta_2^{\alpha} \zeta_{1\gamma})
+ f_+^{\alpha(k-1)\dot\alpha(k-2)\dot\gamma}
(\zeta_1^{\dot\alpha}\zeta_{2\dot\gamma} - \zeta_2^{\dot\alpha}
\zeta_{1\dot\gamma})]
\end{eqnarray*}
Let us stress that the coefficients in this expression must be
$k$-independent. This gives additional restrictions on the
parameters and also serves as a quite non-trivial test for our
calculations.

Thus to construct massive supermultiplets we start with the four
suitable massive superblocks with four free parameters
$\rho_{1,2,3,4}$. Then we require that the commutator of the two
supertransformations on the bosonic fields be closed. In the next
two subsections we apply this scheme to the massive supermultiplets
with half-integer $Y = s-1/2$ and integer $Y=s$ superspins.

\subsection{Half-integer superspin $S-1/2$}

The massive superspin-(s-1/2) supermultiplet contains
$$
\xymatrix {& {[s-\frac12]_{M_1}^{+}} \ar@{-}[dl]_-{\rho_1}
\ar@{-}[dr]^-{\rho_3} &\\
 {[s]_{M}^+} \ar@{-}[dr]_-{\rho_2} & S-1/2 &
{[{s-1}]_{M'}^-}\\
 & {[s-\frac12]_{M_2}^{-}} \ar@{-}[ur]_-{\rho_4} & }
$$
Firstly, we note that four superblocks give the following relations
on the mass parameters:
\begin{align*}
& M^2=M_1(M_1+\lambda) && {M'}^2=M_1(M_1+\lambda)
\\
& M^2=M_2(M_2-\lambda) && {M'}^2=M_2(M_2-\lambda)
\end{align*}
Their solution is
\begin{equation}
M^2={M'}^2=M_1(M_1+\lambda),\qquad M_2=M_1+\lambda
\end{equation}
All the conditions for the closure of the superalgebra are fulfilled
provided:
\begin{equation}
\rho_1{}^2 = \rho_2{}^2 = \rho_3{}^2 = \rho_4{}^2, \qquad
\rho_1\rho_3 = \rho_2\rho_4 \label{ConHI1}
\end{equation}

If the relations (\ref{ConHI1}) are satisfied then the commutators
of the supertransformations on parity-even spin-$s$ $f_+$ and
parity-odd spin-$(s-1)$ $f_-$ fields have the same form:
\begin{eqnarray*}
\frac{1}{i\rho_0{}^2}{[\delta_1,\delta_2]}
f_\pm^{\alpha(k-1)\dot\alpha(k-1)} &=&
\Omega_\pm^{\alpha(k-1)\gamma\dot\alpha(k-2)} (\zeta_1^{\dot\alpha}
\zeta_{2\gamma} - \zeta_2^{\dot\alpha} \zeta_{1\gamma}) +
\Omega_\pm^{\alpha(k-2)\dot\alpha(k-1)\dot\gamma}
(\zeta_1^{\alpha}\zeta_{2\dot\gamma} - \zeta_2^{\alpha}
\zeta_{1\dot\gamma})
\\
 && + \frac{(k-1)a_{k+1}}{2(k+1)}
f_\pm^{\alpha(k-1)\gamma\dot\alpha(k-1)\dot\beta}
(\zeta_{1\dot\beta}\zeta_{2\gamma} - \zeta_{2\dot\beta}
\zeta_{1\gamma})
\\
 && + \frac{a_{k}}{2k(k-1)}
f_\pm^{\alpha(k-2)\dot\alpha(k-2)} (\zeta_1^{\alpha}
\zeta_2^{\dot\alpha} - \zeta_2^{\alpha} \zeta_1^{\dot\alpha})
\\
 && +\lambda[
f_\pm^{\alpha(k-2)\gamma\dot\alpha(k-1)}(\zeta_1^{\alpha}
\zeta_{2\gamma} - \zeta_2^{\alpha} \zeta_{1\gamma}) +
f_\pm^{\alpha(k-1)\dot\alpha(k-2)\dot\gamma}
(\zeta_1^{\dot\alpha}\zeta_{2\dot\gamma} - \zeta_2^{\dot\alpha}
\zeta_{1\dot\gamma})]
\end{eqnarray*}
\begin{eqnarray*}
\frac{1}{i\rho_0{}^2}{[\delta_1,\delta_2]} f_\pm^{\alpha\dot\alpha}
&=& \Omega_\pm^{\alpha\gamma} (\zeta_1^{\dot\alpha} \zeta_{2\gamma}
-\zeta_2^{\dot\alpha} \zeta_{1\gamma}) +
\Omega_\pm^{\dot\alpha\dot\gamma} (\zeta_1^{\alpha}
\zeta_{2\dot\gamma} - \zeta_2^{\alpha} \zeta_{1\dot\gamma})
\\
 && + \frac{a_3}{6} f_\pm^{\alpha\gamma\dot\alpha\dot\beta}
(\zeta_{1\dot\beta} \zeta_{2\gamma} -
\zeta_{2\dot\beta}\zeta_{1\gamma}) - \frac{a_0}{4} A_\pm
(\zeta_1^\alpha \zeta_2^{\dot\alpha} - \zeta_2^\alpha
\zeta_1^{\dot\alpha})
\\
 && + \lambda[ f_\pm^{\gamma\dot\alpha}
(\zeta_1^{\alpha}\zeta_{2\gamma} - \zeta_2^{\alpha} \zeta_{1\gamma})
+ f_\pm^{\alpha\dot\gamma} (\zeta_1^{\dot\alpha} \zeta_{2\dot\gamma}
- \zeta_2^{\dot\alpha} \zeta_{1\dot\gamma})]
\\
\frac{1}{i\rho_0{}^2}{[\delta_1,\delta_2]} A_\pm &=& -
2[e_{\beta\dot\beta} B_\pm^{\alpha\beta}
(\zeta_1^{\dot\beta}\zeta_{2\alpha} - \zeta_2^{\dot\beta}
\zeta_{1\alpha}) + e_{\beta\dot\beta} B_\pm^{\dot\alpha\dot\beta}
(\zeta_1^{\beta}\zeta_{2\dot\alpha} - \zeta_2^{\beta}
\zeta_{1\dot\alpha})]
\\
 && - \frac{a_0}{2} f_\pm^{\alpha\dot\beta}
(\zeta_{1\dot\beta}\zeta_{2\alpha} - \zeta_{2\dot\beta}
\zeta_{1\alpha})
\\
\frac{1}{i\rho_0{}^2}{[\delta_1,\delta_2]} \varphi_\pm &=&
\pi_\pm^{\alpha\dot\alpha} (\zeta_{1\dot\alpha} \zeta_{2\alpha} -
\zeta_{2\dot\alpha} \zeta_{1\alpha})
\end{eqnarray*}
where $a_k$ is determined by (\ref{boson_date}) for spin $s$ and
spin $(s-1)$ respectively and
$$
\rho_0{}^2=\frac{s(2M_1+\lambda)}{2}\rho_1{}^2
$$

\subsection{Integer superspin $S$}

The massive superspin-s supermultiplet contains
$$
\xymatrix {& {[s+\frac12]_{M_1}^{+}} \ar@{-}[dl]_-{\rho_1}
\ar@{-}[dr]^-{\rho_3} &\\
 {[s]_{M}^{\uparrow}} \ar@{-}[dr]_-{\rho_2} & S &
{[{s}]_{M'}^{\downarrow}}\\
 & {[s-\frac12]_{M_2}^{-}} \ar@{-}[ur]_-{\rho_4} & }
$$
First of all, we note that the four superblocks give the following
relations on the mass parameters
\begin{align*}
& M^2=M_1(M_1-\lambda) && {M'}^2=M_1(M_1+\lambda)
\\
& M^2=M_2(M_2-\lambda) && {M'}^2=M_2(M_2+\lambda)
\end{align*}
Their solution is
\begin{equation}
M^2 = M_1(M_1-\lambda), \qquad {M'}^2 = M_1(M_1+\lambda), \qquad M_2
= M_1
\end{equation}
The requirement that the superalgebra be closed again leads to
\begin{equation}
\rho_1{}^2 = \rho_2{}^2 = \rho_3{}^2 = \rho_4{}^2, \qquad
\rho_1\rho_3 = \rho_2\rho_4\label{ConI1}
\end{equation}

If the relations (\ref{ConI1}) are satisfied then commutators of
supertransformations on parity-even and parity-odd bosonic spin-$s$
fields have the same form:
\begin{eqnarray*}
\frac{1}{i\rho_0{}^2}{[\delta_1,\delta_2]}
f_\pm^{\alpha(k-1)\dot\alpha(k-1)} &=&
\Omega_\pm^{\alpha(k-1)\gamma\dot\alpha(k-2)}
(\zeta_1^{\dot\alpha}\zeta_{2\gamma} - \zeta_2^{\dot\alpha}
\zeta_{1\gamma}) + \Omega_\pm^{\alpha(k-2)\dot\alpha(k-1)\dot\gamma}
(\zeta_1^{\alpha}\zeta_{2\dot\gamma} - \zeta_2^{\alpha}
\zeta_{1\dot\gamma})
\\
 && + \frac{(k-1)a_{k+1}}{2(k+1)}
f_\pm^{\alpha(k-1)\gamma\dot\alpha(k-1)\dot\beta}
(\zeta_{1\dot\beta}\zeta_{2\gamma} - \zeta_{2\dot\beta}
\zeta_{1\gamma})
\\
 && + \frac{a_{k}}{2k(k-1)}
f_\pm^{\alpha(k-2)\dot\alpha(k-2)}(\zeta_1^{\alpha}
\zeta_2^{\dot\alpha} - \zeta_2^{\alpha} \zeta_1^{\dot\alpha})
\\
 && +\lambda[
f_\pm^{\alpha(k-2)\gamma\dot\alpha(k-1)}(\zeta_1^{\alpha}
\zeta_{2\gamma} - \zeta_2^{\alpha} \zeta_{1\gamma}) +
f_\pm^{\alpha(k-1)\dot\alpha(k-2)\dot\gamma}
(\zeta_1^{\dot\alpha}\zeta_{2\dot\gamma} - \zeta_2^{\dot\alpha}
\zeta_{1\dot\gamma})]
\\
\frac{1}{i\rho_0{}^2}{[\delta_1,\delta_2]} f_\pm^{\alpha\dot\alpha}
&=& \Omega_\pm^{\alpha\gamma} (\zeta_1^{\dot\alpha} \zeta_{2\gamma}
-\zeta_2^{\dot\alpha} \zeta_{1\gamma}) +
\Omega_\pm^{\dot\alpha\dot\gamma} (\zeta_1^{\alpha}
\zeta_{2\dot\gamma} - \zeta_2^{\alpha} \zeta_{1\dot\gamma})
\\
 && + \frac{a_3}{6} f_\pm^{\alpha\gamma\dot\alpha\dot\beta}
(\zeta_{1\dot\beta} \zeta_{2\gamma} -
\zeta_{2\dot\beta}\zeta_{1\gamma}) -\frac{a_0}{4} A_\pm
(\zeta_1^\alpha \zeta_2^{\dot\alpha} - \zeta_2^\alpha
\zeta_1^{\dot\alpha})
\\
 && +\lambda[ f_\pm^{\gamma\dot\alpha}
(\zeta_1^{\alpha}\zeta_{2\gamma} - \zeta_2^{\alpha} \zeta_{1\gamma})
+ f_\pm^{\alpha\dot\gamma} (\zeta_1^{\dot\alpha} \zeta_{2\dot\gamma}
- \zeta_2^{\dot\alpha} \zeta_{1\dot\gamma})]
\\
\frac{1}{i\rho_0{}^2}{[\delta_1,\delta_2]} A_\pm &=& - 2
[e_{\beta\dot\beta} B_\pm^{\alpha\beta}
(\zeta_1^{\dot\beta}\zeta_{2\alpha} - \zeta_2^{\dot\beta}
\zeta_{1\alpha}) + e_{\beta\dot\beta} B_\pm^{\dot\alpha\dot\beta}
(\zeta_1^{\beta}\zeta_{2\dot\alpha} - \zeta_2^{\beta}
\zeta_{1\dot\alpha})]
\\
 && -\frac{a_0}{2} f_\pm^{\alpha\dot\beta}
(\zeta_{1\dot\beta}\zeta_{2\alpha} - \zeta_{2\dot\beta}
\zeta_{1\alpha})
\\
\frac{1}{i\rho_0{}^2}{[\delta_1,\delta_2]} \varphi_\pm &=&
\pi_\pm^{\alpha\dot\alpha} (\zeta_{1\dot\alpha} \zeta_{2\alpha} -
\zeta_{2\dot\alpha} \zeta_{1\alpha})
\end{eqnarray*}
where $a_k$ is determined by (\ref{boson_date}) for spin $s$ and
$$
\rho_0{}^2=\frac{(2s+1)M_1}{2}\rho_1{}^2
$$

\section{Summary}

In this paper we have developed the component Lagrangian description
of massive on-shell $N=1$ supermultiplets with arbitrary
(half)integer superspin in four dimensional Anti de Sitter space
($AdS_4$). The derivation is based on supersymmetric generalization
of frame-like gauge invariant formulation of massive higher spin
fields where massive supermultiplets are described by an appropriate
set of massless ones. We show that $N=1$ massive supermultiplets can
be constructed as a combination of four massive superblocks, each
containing one massive boson and one massive fermion. As a result,
we have derived both the supertransformations for the components of
the one-shell supermultiplets and the corresponding invariant
Lagrangians. Thus, the component Lagrangian formulation of the $N=1$
supersymmetric free massive higher spin field theory on the $AdS_4$
space can be considered complete.

Let us briefly discuss the possible further generalizations of the
results obtained. As we already pointed out, the problem of
off-shell supersymmetric massive higher spin theory remains open in
general. Only a few examples of such a theory with concrete
superspins \cite{BG1}, \cite{BG2}, \cite{BG3} in flat space have
been developed. There are no known examples in the $AdS$ space.
There are two possible approaches to study this general problem. One
option is to start with on-shell theory and try to find the
necessary auxiliary fields closing the superalgebra on the base of
Noether's procedure. Another approach can be based on the use of the
superfield techniques from the very beginning. At present,
realization of both these approaches seems unclear and will require
the development of new methods. Besides, the interesting
generalizations of the results obtained can be constructing the
partially massless $N=1$ supermultiplets and finding at least
on-shell component Lagrangian description for $N$-extended massive
supermultiplets in flat and $AdS$ spaces. We hope to attack these
problems in the forthcoming works.

\section*{Acknowledgments}
The authors are grateful to S.M. Kuzenko for useful comments. I.L.B and T.V.S are thankful to the
RFBR grant, project No. 18-02-00153-a for partial support. Their research was also supported
in parts by Russian Ministry of Science and High Education, project No.
3.1386.2017.

\appendix

\section{Identities}
In this appendix we present explicit expression of identities for
massive higher spin superblocks.

\subsection{Auxiliary bosonic fields}

Identities that correspond to (\ref{SchemIden1})
\begin{eqnarray*}
0 &=& {\cal F}_{\alpha(k-1)\beta\dot\alpha(k-1)} e^\beta{}_{\dot\beta}
\Omega^{\alpha(k-2)\dot\alpha(k-1)\dot\beta} \zeta^{\alpha}
+ \Phi_{\alpha(k-2)\beta\gamma\dot\alpha(k-1)} e^\beta{}_{\dot\beta}
{\cal R}^{\alpha(k-2)\dot\alpha(k-1)\dot\beta} \zeta^{\gamma}
\\
 && - 2{d_k}[(k+1) E^{\dot\beta}{}_{\dot\gamma}
\Psi^{\alpha(k-2)\gamma\dot\alpha(k-1)\dot\gamma}
\Omega_{\alpha(k-2)\dot\alpha(k-1)\dot\beta} \zeta_{\gamma}
\\
 && - (k-2) E_\beta{}^\alpha
\Psi^{\alpha(k-3)\beta\gamma\dot\alpha(k)}
\Omega_{\alpha(k-2)\dot\alpha(k)} \zeta_{\gamma} - E_\beta{}^\gamma
\Psi^{\alpha(k-2)\beta\dot\alpha(k)}
\Omega_{\alpha(k-2)\dot\alpha(k)} \zeta_{\gamma} ]
\\
 && - \frac{b_k}{2k} [(k+1)E^\beta{}_\delta
\Phi_{\alpha(k-2)\beta\gamma\dot\alpha(k-1)}
f^{\alpha(k-2)\delta\dot\alpha(k-1)} - (k-1)
E_{\dot\beta}{}^{\dot\alpha}
\Phi_{\alpha(k-1)\gamma\dot\alpha(k-1)}f^{\alpha(k-1)\dot\alpha(k-2)\dot\beta}  ]\zeta^{\gamma}
\\
 && + \lambda [E^{\alpha\beta} \Phi_{\alpha(k-1)\beta\dot\alpha(k-1)}
\Omega^{\alpha(k-2)\dot\alpha(k-1)\dot\beta} \zeta_{\dot\beta}]
\\
 && - c_{k+1}[-E_{\beta\gamma}
\Phi^{\alpha(k-2)\delta\beta\gamma\dot\alpha(k)}
\Omega_{\alpha(k-2)\dot\alpha(k)} \zeta_{\delta}]
\\
 && - \frac{(k-1)c_k}{(k-1)(k+1)}
[kE^{\dot\beta\dot\alpha}\Phi^{\alpha(k-2)\delta\dot\alpha(k-2)}
\Omega_{\alpha(k-2)\dot\alpha(k-1)\dot\beta} \zeta_{\delta} +
E^{\dot\beta\dot\alpha} \Phi^{\alpha(k-2)\beta\dot\alpha(k-2)}
\Omega_{\alpha(k-2)\dot\alpha(k-1)\dot\beta} \zeta_{\beta}]
\\
 && - \frac{a_{k+1}}{2} [- E_{\dot\beta\dot\gamma}
\Phi_{\alpha(k-1)\delta\dot\alpha(k-1)}
\Omega^{\alpha(k-1)\dot\alpha(k-1)\dot\beta\dot\gamma}
]\zeta^{\delta}\\
 && - \frac{(k-2)a_k}{2k(k+1)} [(k+1)E^{\beta\alpha}
\Phi_{\alpha(k-2)\beta\gamma\dot\alpha(k-1)}
\Omega^{\alpha(k-3)\dot\alpha(k-1)}] \zeta^{\gamma}
\end{eqnarray*}
\begin{eqnarray*}
0 &=& {\cal F}_{\alpha(k-2)\gamma\dot\alpha(k-2)}
e^\gamma{}_{\dot\gamma}
\Omega^{\alpha(k-2)\dot\alpha(k-2)\dot\gamma\dot\beta}
\zeta_{\dot\beta} + \Phi_{\alpha(k-2)\gamma\dot\alpha(k-2)}
e^\gamma{}_{\dot\gamma}
{\cal R}^{\alpha(k-2)\dot\alpha(k-2)\dot\gamma\dot\beta}
\zeta_{\dot\beta}
\\
 && + 2{d_{k-1}}[k E^{\dot\gamma}{}_{\dot\delta}
\Psi^{\alpha(k-2)\dot\alpha(k-2)\dot\delta}
\Omega_{\alpha(k-2)\dot\alpha(k-2)\dot\gamma\dot\beta}
\zeta^{\dot\beta}
\\
 && - (k-2)E_\gamma{}^\beta \Psi^{\alpha(k-3)\gamma\dot\alpha(k-1)}
\Omega_{\alpha(k-3)\beta\dot\alpha(k-1)\dot\beta} \zeta^{\dot\beta}]
\\
 && - \frac{b_{k}}{2k} [-
(k-2)E_{\dot\gamma}{}^{\dot\alpha}\Phi_{\alpha(k-1)\dot\alpha(k-2)}
f^{\alpha(k-1)\dot\alpha(k-3)\dot\beta\dot\gamma} + (k+1)
E^\gamma{}_\beta \Phi_{\alpha(k-2)\gamma\dot\alpha(k-2)}
f^{\alpha(k-2)\beta\dot\alpha(k-2)\dot\beta})
\\
 && - E_{\dot\gamma}{}^{\dot\beta} \Phi_{\alpha(k-1)\dot\alpha(k-2)}
f^{\alpha(k-1)\dot\alpha(k-2)\dot\gamma}] \zeta_{\dot\beta}
\\
 && + \lambda [E_{\dot\beta\dot\gamma}
\Phi_{\alpha(k-2)\gamma\dot\alpha(k-2)}
\Omega^{\alpha(k-2)\dot\alpha(k-2)\dot\gamma\dot\beta} \zeta^\gamma]
\\
 && + c_{k}[-E_{\gamma\delta}
\Phi^{\alpha(k-2)\gamma\delta\dot\alpha(k-1)}
\Omega_{\alpha(k-2)\dot\alpha(k-1)\dot\beta} \zeta^{\dot\beta}]
\\
 && + \frac{(k-2)c_{k-1}}{k(k-2)}
[kE^{\dot\gamma\dot\alpha}\Phi^{\alpha(k-2)\dot\alpha(k-3)}
\Omega_{\alpha(k-2)\dot\alpha(k-2)\dot\gamma\dot\beta}
\zeta^{\dot\beta}]
\\
 && - \frac{a_{k+1}}{2} [- E_{\dot\gamma\dot\delta}
\Phi_{\alpha(k-1)\dot\alpha(k-2)}
\Omega^{\alpha(k-1)\dot\alpha(k-2)\dot\gamma\dot\beta\dot\delta}]
\zeta_{\dot\beta}
\\
 && - \frac{(k-2)a_k}{2k(k+1)} [(k+1)E^{\gamma\alpha}
\Phi_{\alpha(k-2)\gamma\dot\alpha(k-2)}
\Omega^{\alpha(k-3)\dot\alpha(k-2)\dot\beta}] \zeta_{\dot\beta}
\end{eqnarray*}
\begin{eqnarray*}
0 &=& {\cal F}_{\alpha\beta\dot\alpha}e^\beta{}_{\dot\beta}
\Omega^{\dot\alpha\dot\beta}\zeta^{\alpha}
+ \Phi_{\beta\gamma\dot\alpha} e^\beta{}_{\dot\beta}
{\cal R}^{\dot\alpha\dot\beta}\zeta^{\gamma}
\\
 && - 2d_2[3E^{\dot\beta}{}_{\dot\gamma}
\Psi^{\gamma\dot\alpha\dot\gamma} \Omega_{\dot\alpha\dot\beta}
\zeta_{\gamma} - E_\beta{}^\gamma \Psi^{\beta\dot\alpha(2)}
\Omega_{\dot\alpha(2)} \zeta_{\gamma} ]
\\
 && - \frac{b_2}{4}
[3E^\beta{}_\delta\Phi_{\beta\gamma\dot\alpha}f^{\delta\dot\alpha} -
E_{\dot\beta}{}^{\dot\alpha}
\Phi_{\alpha\gamma\dot\alpha} f^{\alpha\dot\beta}]\zeta^{\gamma}
+\lambda [E^{\alpha\beta} \Phi_{\alpha\beta\dot\alpha}
\Omega^{\dot\alpha\dot\beta} \zeta_{\dot\beta}]
\\
 && - c_3[-E_{\beta\gamma} \Phi^{\delta\beta\gamma\dot\alpha(2)}
\Omega_{\dot\alpha(2)}\zeta_{\delta}] -\ frac{c_2}{3}
[2E^{\dot\beta\dot\alpha} \Phi^{\delta}
\Omega_{\dot\alpha\dot\beta}\zeta_{\delta} + E^{\dot\beta\dot\alpha}
\Phi^{\beta}
\Omega_{\dot\alpha\dot\beta} \zeta_{\beta}]
\\
 && - \frac{a_{3}}{2} [- E_{\dot\beta\dot\gamma}
\Phi_{\alpha\delta\dot\alpha}
\Omega^{\alpha\dot\alpha\dot\beta\dot\gamma}] \zeta^{\delta} -
[\frac{a_0}{4} (4E^\beta{}_{\dot\gamma} \Phi_{\alpha\beta\dot\alpha}
B^{\dot\alpha\dot\gamma}) - \frac{a_0\tilde{a}_0}{8}
E^{\beta\dot\alpha} \Phi_{\alpha\beta\dot\alpha}\varphi]
\zeta^{\alpha}
\\
0 &=& {\cal F}_{\alpha}e^\alpha{}_{\dot\alpha}
\Omega^{\dot\alpha\dot\beta} \zeta_{\dot\beta} + \Phi_{\alpha}
e^\alpha{}_{\dot\alpha} {\cal R}^{\dot\alpha\dot\beta}
\zeta_{\dot\beta}
\\
 && + [4{d_{1}}E_{\dot\alpha\dot\beta} \Phi^{\dot\beta}
\Omega^{\dot\alpha\dot\gamma} \zeta_{\dot\gamma} - c_{2}
E_{\alpha\beta} \Phi^{\alpha\beta\dot\beta}
\Omega_{\dot\beta\dot\gamma} \zeta^{\dot\gamma} + c_0
E_{\beta\dot\alpha} \phi^\beta \Omega^{\dot\alpha\dot\gamma}
\zeta_{\dot\gamma}]
\\
 && - [-\frac{a_3}{2} E_{\dot\alpha\dot\beta} \Phi_{\beta}
\Omega^{\beta\dot\alpha\dot\gamma\dot\beta} + \frac{b_2}{4}
(3E^\alpha{}_\beta \Phi_{\alpha} f^{\beta\dot\gamma} -
E_{\dot\alpha}{}^{\dot\gamma} \Phi_{\beta} f^{\beta\dot\alpha})
\\
 && + {a_0}E^\alpha{}_{\dot\beta} \Phi_{\alpha}
B^{\dot\gamma\dot\beta} - \frac{a_0\tilde{a}_0}{8}
E^{\alpha\dot\gamma} \Phi_{\alpha}\varphi] \zeta_{\dot\gamma} +
\lambda E_{\dot\beta\dot\alpha} \Phi_{\alpha}
\Omega^{\dot\alpha\dot\beta} \zeta^\alpha
\end{eqnarray*}
\begin{eqnarray*}
0 &=& E_{\dot\alpha\dot\beta} {\cal F}_{\alpha}
B^{\dot\alpha\dot\beta} \zeta^{\alpha} -
E_{\dot\alpha\dot\beta}\Phi_{\alpha} {\cal C}^{\dot\alpha\dot\beta}
\zeta^{\alpha} + [4{d_{1}} E^\alpha{}_{\dot\gamma} \Phi_{\dot\alpha}
B^{\dot\alpha\dot\gamma} \zeta_{\alpha} -
2c_2E_\beta{}_{\dot\gamma}\Phi^{\alpha\beta\dot\beta}
B_{\dot\beta}{}^{\dot\gamma}\zeta_{\alpha}]
\\
 && + [-\frac{a_0}{2} E_{\dot\alpha\dot\beta}
\Phi_{\alpha}\Omega^{\dot\alpha\dot\beta} - \frac{\tilde{a}_0}{4}
E_{\beta\dot\beta}\Phi_{\alpha} \pi^{\beta\dot\beta}]\zeta^{\alpha} -
2\lambda E^\alpha{}_{\dot\beta} \Phi_{\alpha}B^{\dot\alpha\dot\beta}
\zeta_{\dot\alpha}
\\
0 &=& {\cal C}_\alpha E^\alpha{}_{\dot\alpha}
B^{\dot\alpha\dot\beta}\zeta_{\dot\beta} - \phi_\alpha
E^\alpha{}_{\dot\alpha}
{\cal C}^{\dot\alpha\dot\beta} \zeta_{\dot\beta}
\\
 && + [c_0E_\alpha{}_{\dot\alpha} \Phi^{\alpha}
B^{\dot\alpha\dot\gamma} \zeta_{\dot\gamma} + 4d_1E\phi^{\dot\alpha}
B^{\dot\alpha\dot\gamma} \zeta_{\dot\gamma}]
+[-\frac{a_0}{2} E^\alpha{}_{\dot\alpha} \phi_\alpha
\Omega^{\dot\alpha\dot\beta} + \frac{\tilde{a}_0}{8} E\phi_\beta
\pi^{\beta\dot\beta}] \zeta_{\dot\beta}
\\
0 &=& - E^\alpha{}_{\dot\alpha} {\cal C}_\alpha
\pi^{\beta\dot\alpha} \zeta_{\beta} + \frac12
E_\gamma{}_{\dot\alpha}\phi^\beta {\cal
C}^{\gamma\dot\alpha}\zeta_{\beta}
-[-c_0E_\alpha{}_{\dot\alpha} \Phi^{\alpha}
\pi^{\beta\dot\alpha}\zeta_{\beta} -4d_1E\phi_{\dot\alpha}
\pi^{\beta\dot\alpha}\zeta_{\beta}]
\\
 && + [\frac{a_0\tilde{a}_0}{24} E^\alpha{}_{\dot\alpha} \phi_\alpha
f^{\beta\dot\alpha} - \frac{\tilde{a}_0}{12} (-2E\phi_\gamma
B^{\beta\gamma}) - \frac{a_0{}^2}{4} \phi^\beta \varphi]
\zeta_{\beta} + \lambda E\phi_\beta \pi^{\beta\dot\alpha}
\zeta_{\dot\alpha}
\end{eqnarray*}

\subsection{Physical bosonic fields}

Identities that correspond to (\ref{SchemIden2})
\begin{eqnarray*}
0 &=& {\cal F}_{\alpha(k-1)\beta\dot\alpha(k-1)} e^\beta{}_{\dot\beta}
f^{\alpha(k-1)\dot\alpha(k-1)} \zeta^{\dot\beta}
\\
 && - 2{d_k}[(k+1)E^{\dot\beta}{}_{\dot\gamma}
\Psi^{\alpha(k-1)\dot\alpha(k-1)\dot\gamma}
f_{\alpha(k-1)\dot\alpha(k-1)} \zeta_{\dot\beta}
\\
 && - (k-1)E_\beta{}^\alpha
\Psi^{\alpha(k-2)\beta\dot\alpha(k-1)\dot\beta}
f_{\alpha(k-1)\dot\alpha(k-1)}\zeta_{\dot\beta} ]
\\
 && - [ (k-1)(- E_{\dot\beta}{}^{\dot\alpha}
\Phi_{\alpha(k)\dot\alpha(k-1)}
\Omega^{\alpha(k)\dot\alpha(k-2)} + E^\beta{}_\gamma
\Phi_{\alpha(k-1)\beta\dot\alpha(k-2)\dot\beta}
\Omega^{\alpha(k-1)\gamma\dot\alpha(k-2)}) \zeta^{\dot\beta}
\\
 && + (k-1)E^{\beta\alpha}
\Phi_{\alpha(k-1)\beta\dot\alpha(k-1)}\Omega^{\alpha(k-2)\dot\alpha(k-1)\dot\gamma} \zeta_{\dot\gamma}]
\\
 && - 2\lambda E_\gamma{}^\beta
\Phi_{\alpha(k-1)\beta\dot\alpha(k-1)}f^{\alpha(k-1)\dot\alpha(k-1)}
\zeta^{\gamma}
\\
 && - c_{k+1}[-E_{\beta\gamma}
\Phi^{\alpha(k-1)\beta\gamma\dot\alpha(k-1)\dot\beta}
f_{\alpha(k-1)\dot\alpha(k-1)} \zeta_{\dot\beta}]
\\
 && - \frac{(k-1)c_k}{(k-1)(k+1)}
[(k+1)E^{\dot\beta\dot\alpha}\Phi^{\alpha(k-1)\dot\alpha(k-2)}
f_{\alpha(k-1)\dot\alpha(k-1)}\zeta_{\dot\beta}
\\
 && - (k-1)E_\beta{}^\alpha \Phi^{\alpha(k-2)\beta\dot\alpha(k-2)}
f_{\alpha(k-1)\dot\alpha(k-2)\dot\beta} \zeta^{\dot\beta}]
\\
 && - \frac{(k-1)a_{k+1}}{2(k+1)}[-
E_{\dot\beta\dot\gamma}\Phi_{\alpha(k)\dot\alpha(k-1)}
f^{\alpha(k)\dot\alpha(k-1)\dot\gamma}\zeta^{\dot\beta} +
E^\beta{}_\gamma
\Phi_{\alpha(k-1)\beta\dot\alpha(k-1)}
f^{\alpha(k-1)\gamma\dot\alpha(k-1)\dot\gamma} \zeta_{\dot\gamma} ]
\\
 && - \frac{(k-1)^2a_k}{2k(k-1)}[E^{\beta\alpha}
\Phi_{\alpha(k-1)\beta\dot\alpha(k-2)\dot\beta}
f^{\alpha(k-2)\dot\alpha(k-2)}] \zeta^{\dot\beta}
\end{eqnarray*}
\begin{eqnarray*}
0 &=& {\cal F}_{\alpha(k-1)\beta\dot\alpha(k-2)\dot\gamma}
e^\beta{}_{\dot\beta}
f^{\alpha(k-1)\dot\alpha(k-2)\dot\beta}\zeta^{\dot\gamma}
\\
 && - 2{d_k}[(k+1)E^{\dot\beta}{}_{\dot\delta}
\Psi^{\alpha(k-1)\dot\alpha(k-2)\dot\gamma\dot\delta}
f_{\alpha(k-1)\dot\alpha(k-2)\dot\beta} \zeta_{\dot\gamma}
\\
 && - (k-1)E_\beta{}^\alpha
\Psi^{\alpha(k-2)\beta\dot\alpha(k-1)\dot\gamma}
f_{\alpha(k-1)\dot\alpha(k-1)} \zeta_{\dot\gamma}]
\\
 && - [kE^\beta{}_\delta
\Phi_{\alpha(k-1)\beta\dot\alpha(k-2)\dot\gamma}
\Omega^{\alpha(k-1)\delta\dot\alpha(k-2)} -(k-2)
E_{\dot\beta}{}^{\dot\alpha}
\Phi_{\alpha(k)\dot\alpha(k-2)\dot\gamma}\Omega^{\alpha(k)\dot\alpha(k-3)\dot\beta} ]\zeta^{\dot\gamma}
\\
 && + \lambda[ E^{\dot\gamma}{}_{\dot\beta}
\Phi_{\alpha(k-1)\gamma\dot\alpha(k-2)\dot\gamma}
f^{\alpha(k-1)\dot\alpha(k-2)\dot\beta} \zeta^{\gamma} -
E_\gamma{}^\beta \Phi_{\alpha(k-1)\beta\dot\alpha(k-1)}
f^{\alpha(k-1)\dot\alpha(k-1)} \zeta^{\gamma}]
\\
 && - c_{k+1}[-E_{\beta\delta}
\Phi^{\alpha(k-1)\beta\delta\dot\alpha(k-1)\dot\gamma}
f_{\alpha(k-1)\dot\alpha(k-1)} \zeta_{\dot\gamma}]
\\
 && - \frac{c_k}{(k-1)(k+1)}[(k-2)(k+1)
E^{\dot\beta\dot\alpha}\Phi^{\alpha(k-1)\dot\alpha(k-3)\dot\gamma}
f_{\alpha(k-1)\dot\alpha(k-2)\dot\beta} \zeta_{\dot\gamma}
\\
 && + (k+1) E^{\dot\beta\dot\gamma} \Phi^{\alpha(k-1)\dot\alpha(k-2)}
f_{\alpha(k-1)\dot\alpha(k-2)\dot\beta} \zeta_{\dot\gamma} + (k-1)
E_\beta{}^\alpha \Phi^{\alpha(k-2)\beta\dot\alpha(k-2)}
f_{\alpha(k-1)\dot\alpha(k-2)\dot\beta} \zeta^{\dot\beta}]
\\
 && - \frac{(k-1)a_{k+1}}{2(k+1)} [-
E_{\dot\beta\dot\gamma}\Phi_{\alpha(k)\dot\alpha(k-2)\dot\delta}
f^{\alpha(k)\dot\alpha(k-2)\dot\beta\dot\gamma} ]\zeta^{\dot\delta}
\\
 && - \frac{(k-1)a_k}{2k(k-1)} [kE^{\beta\alpha}
\Phi_{\alpha(k-1)\beta\dot\alpha(k-2)\dot\gamma}
f^{\alpha(k-2)\dot\alpha(k-2)}] \zeta^{\dot\gamma}
\end{eqnarray*}
\begin{eqnarray*}
0 &=& {\cal F}_{\alpha(k-2)\gamma\dot\alpha(k-2)}
e^\gamma{}_{\dot\gamma}
f^{\alpha(k-2)\beta\dot\alpha(k-2)\dot\gamma}\zeta_\beta
\\
 && + 2{d_{k-1}}[k E^{\dot\gamma}{}_{\dot\beta}
\Psi^{\alpha(k-2)\dot\alpha(k-2)\dot\beta}
f_{\alpha(k-2)\beta\dot\alpha(k-2)\dot\gamma} \zeta^\beta
\\
 && - (k-2)E_\gamma{}^\alpha \Psi^{\alpha(k-3)\gamma\dot\alpha(k-1)}
f_{\alpha(k-2)\beta\dot\alpha(k-1)} \zeta^\beta]
\\
 && - [-(k-2) E_{\dot\gamma}{}^{\dot\alpha}
\Phi_{\alpha(k-1)\dot\alpha(k-2)}
\Omega^{\alpha(k-1)\beta\dot\alpha(k-3)\dot\gamma} \zeta_\beta
\\
 && + kE^\gamma{}_\delta
\Phi_{\alpha(k-2)\gamma\dot\alpha(k-2)}\Omega^{\alpha(k-2)\beta\delta\dot\alpha(k-2)} \zeta_\beta +
E_{\dot\gamma\dot\beta} \Phi_{\alpha(k-2)\gamma\dot\alpha(k-2)}
\Omega^{\alpha(k-2)\dot\alpha(k-2)\dot\gamma\dot\beta} \zeta^\gamma]
\\
 && + \lambda[ E_{\dot\beta\dot\gamma}
\Phi_{\alpha(k-1)\dot\alpha(k-2)}
f^{\alpha(k-1)\dot\alpha(k-2)\dot\gamma} \zeta^{\dot\beta} +
E_\beta{}^\gamma \Phi_{\alpha(k-2)\gamma\dot\alpha(k-2)}
f^{\alpha(k-2)\beta\dot\alpha(k-2)\dot\gamma} \zeta_{\dot\gamma} ]
\\
 && + c_{k}[-E_{\gamma\delta}
\Phi^{\alpha(k-2)\gamma\delta\dot\alpha(k-1)}
f_{\alpha(k-2)\beta\dot\alpha(k-1)} \zeta^\beta]
\\
 && +
\frac{(k-2)c_{k-1}}{k(k-2)}[kE^{\dot\gamma\dot\alpha}\Phi^{\alpha(k-2)\dot\alpha(k-3)}
f_{\alpha(k-2)\beta\dot\alpha(k-2)\dot\gamma} \zeta^\beta]
\\
 && - \frac{(k-1)a_{k+1}}{2(k+1)} [-
E_{\dot\gamma\dot\delta}\Phi_{\alpha(k-1)\dot\alpha(k-2)}
f^{\alpha(k-1)\beta\dot\alpha(k-2)\dot\gamma\dot\delta}] \zeta_\beta
\\
 && - \frac{a_k}{2k(k-1)} [k(k-2)E^{\gamma\alpha}
\Phi_{\alpha(k-2)\gamma\dot\alpha(k-2)}
f^{\alpha(k-3)\beta\dot\alpha(k-2)} \zeta_\beta
\\
 && + (k-2)E_{\dot\gamma}{}^{\dot\alpha}
\Phi_{\alpha(k-2)\gamma\dot\alpha(k-2)}
f^{\alpha(k-2)\dot\alpha(k-3)\dot\gamma} \zeta^\gamma +
kE^{\gamma\beta} \Phi_{\alpha(k-2)\gamma\dot\alpha(k-2)}
f^{\alpha(k-2)\dot\alpha(k-2)} \zeta_\beta]
\end{eqnarray*}
\begin{eqnarray*}
0 &=& {\cal F}_{\alpha\beta\dot\alpha} e^\beta{}_{\dot\beta}
f^{\alpha\dot\alpha} \zeta^{\dot\beta}
\\
 && - 2{d_2}[3E^{\dot\beta}{}_{\dot\gamma}
\Psi^{\alpha\dot\alpha\dot\gamma} f_{\alpha\dot\alpha}
\zeta_{\dot\beta} - E_\beta{}^\alpha \Psi^{\beta\dot\alpha\dot\beta}
f_{\alpha\dot\alpha} \zeta_{\dot\beta} ]
\\
 && - [ (- E_{\dot\beta}{}^{\dot\alpha} \Phi_{\alpha(2)\dot\alpha}
\Omega^{\alpha(2)} + E^\beta{}_\gamma \Phi_{\alpha\beta\dot\beta}
\Omega^{\alpha\gamma}) \zeta^{\dot\beta} + E^{\beta\alpha}
\Phi_{\alpha\beta\dot\alpha} \Omega^{\dot\alpha\dot\gamma}
\zeta_{\dot\gamma}]
\\
 && - 2\lambda E_\gamma{}^\beta \Phi_{\alpha\beta\dot\alpha}
f^{\alpha\dot\alpha} \zeta^{\gamma}
\\
 && - c_{3}[-E_{\beta\gamma}
\Phi^{\alpha\beta\gamma\dot\alpha\dot\beta}
f_{\alpha\dot\alpha} \zeta_{\dot\beta}] - \frac{c_2}{3}
[3E^{\dot\beta\dot\alpha} \Phi^{\alpha}
f_{\alpha\dot\alpha}\zeta_{\dot\beta} - E_\beta{}^\alpha \Phi^{\beta}
f_{\alpha\dot\beta} \zeta^{\dot\beta}]
\\
 && - \frac{a_3}{6}[- E_{\dot\beta\dot\gamma}
\Phi_{\alpha(2)\dot\alpha}
f^{\alpha(2)\dot\alpha\dot\gamma}\zeta^{\dot\beta} + E^\beta{}_\gamma
\Phi_{\alpha\beta\dot\alpha}
f^{\alpha\gamma\dot\alpha\dot\gamma} \zeta_{\dot\gamma} ]+
\frac{a_0}{4} [E^{\beta\alpha} \Phi_{\alpha\beta\dot\beta} A]
\zeta^{\dot\beta}
\\
0 &=& {\cal F}_{\alpha\beta\dot\gamma} e^\beta{}_{\dot\beta}
f^{\alpha\dot\beta} \zeta^{\dot\gamma}
\\
 && - 2d_2[3E^{\dot\beta}{}_{\dot\delta}
\Psi^{\alpha\dot\gamma\dot\delta} f_{\alpha\dot\beta}
\zeta_{\dot\gamma} - E_\beta{}^\alpha \Psi^{\beta\dot\alpha\dot\gamma}
f_{\alpha\dot\alpha} \zeta_{\dot\gamma}]
\\
 && - [2E^\beta{}_\delta \Phi_{\alpha\beta\dot\gamma}
\Omega^{\alpha\delta} ]\zeta^{\dot\gamma} + \lambda[
E^{\dot\gamma}{}_{\dot\beta} \Phi_{\alpha\gamma\dot\gamma}
f^{\alpha\dot\beta} \zeta^{\gamma} - E_\gamma{}^\beta
\Phi_{\alpha\beta\dot\alpha} f^{\alpha\dot\alpha} \zeta^{\gamma}]
\\
 && - c_{3}[-E_{\beta\delta}
\Phi^{\alpha\beta\delta\dot\alpha\dot\gamma}
f_{\alpha\dot\alpha}\zeta_{\dot\gamma}] - \frac{c_2}{3}[3
E^{\dot\beta\dot\gamma}\Phi^{\alpha} f_{\alpha\dot\beta}
\zeta_{\dot\gamma} +
E_\beta{}^\alpha \Phi^{\beta} f_{\alpha\dot\beta} \zeta^{\dot\beta}]
\\
 && - \frac{a_{3}}{6} [- E_{\dot\beta\dot\gamma}
\Phi_{\alpha(2)\dot\delta} f^{\alpha(2)\dot\beta\dot\gamma}
]\zeta^{\dot\delta} + \frac{a_0}{4} [2E^{\beta\alpha}
\Phi_{\alpha\beta\dot\gamma} A] \zeta^{\dot\gamma}
\\
0 &=& {\cal F}_{\alpha}e^\alpha{}_{\dot\alpha}
f^{\beta\dot\alpha}\zeta_\beta + [4{d_{1}} E_{\dot\alpha\dot\beta}
\Phi^{\dot\beta}f^{\gamma\dot\alpha} \zeta_\gamma -c_2E_{\alpha\beta}
\Phi^{\alpha\beta\dot\beta} f_{\gamma\dot\beta} \zeta^\gamma
+ c_0E_{\beta\dot\alpha} \phi^\beta f^{\gamma\dot\alpha}
\zeta_\gamma]\\
 && - [2E^\alpha{}_\beta \Phi_{\alpha} \Omega^{\gamma\beta} -
E_{\dot\alpha\dot\beta} \Phi^{\gamma} \Omega^{\dot\alpha\dot\beta}
- \frac{a_{3}}{6} E_{\dot\alpha\dot\beta} \Phi_{\beta}
f^{\gamma\beta\dot\alpha\dot\beta} -\frac{a_0}{2}
E^{\alpha\gamma}\Phi_{\alpha}A] \zeta_\gamma
\\
 && + \lambda( E_{\dot\beta\dot\alpha} \Phi_{\beta}
f^{\beta\dot\alpha} \zeta^{\dot\beta} + E_\beta{}^\alpha \Phi_\alpha
f^{\beta\dot\alpha} \zeta_{\dot\alpha})
\end{eqnarray*}
\begin{eqnarray*}
0 &=& - e^\alpha{}_{\dot\alpha} {\cal F}_{\alpha} A
\zeta^{\dot\alpha} - [4{d_{1}} E_{\dot\alpha\dot\beta}
\Phi^{\dot\beta}A \zeta^{\dot\alpha} + c_2 E_{\alpha\beta}
\Phi^{\alpha\beta\dot\beta} A \zeta_{\dot\beta} + c_0
E_{\beta\dot\alpha} \phi^\beta A \zeta^{\dot\alpha}]
\\
 && - [ 4(E_{\alpha\dot\beta} \Phi_{\alpha}B^{\alpha(2)}
\zeta^{\dot\beta} + E^\beta{}_{\dot\alpha} \Phi_{\beta}
B^{\dot\alpha(2)} \zeta_{\dot\alpha})
-\frac{a_0}{2} ( E_{\dot\alpha\dot\beta} \Phi_{\alpha}
f^{\alpha\dot\alpha} \zeta^{\dot\beta} - E_\alpha{}^\beta \Phi_{\beta}
f^{\alpha\dot\alpha} \zeta_{\dot\alpha})]
\\
 && + 2\lambda E^\alpha{}_\beta \Phi_{\alpha} A \zeta^\beta
\\
0 &=& - E^\alpha{}_{\dot\alpha} {\cal C}_\alpha \varphi
\zeta^{\dot\alpha} - [-c_0E_\alpha{}_{\dot\alpha} \Phi^{\alpha}
\varphi \zeta^{\dot\alpha} - 4d_1E \phi_{\dot\alpha}
\varphi\zeta^{\dot\alpha}]
\\
 && + [-E\phi_\alpha \pi^{\alpha\dot\alpha} \zeta_{\dot\alpha}
- \frac{\tilde{a}_0}{12} E^\beta{}_{\dot\beta} \phi_\beta A
\zeta^{\dot\beta}] - 2\lambda E\phi_\alpha \varphi \zeta^\alpha
\end{eqnarray*}

\end{document}